\documentstyle[twocolumn]{jpsj} 
\input epsf.tex
\newcommand{\lsim}
 {\ \raise.35ex\hbox{$<$}\kern-0.75em\lower.5ex\hbox{$\sim$}\ }
\newcommand{\gsim}
 {\ \raise.35ex\hbox{$>$}\kern-0.75em\lower.5ex\hbox{$\sim$}\ }
\def\runtitle{
Filling and Bandwidth Control Mott Transitions:
Grand-Canonical Path-Integral Renormalization Group Approach
}
\def\runauthor{Shinji {\sc Watanabe} and Masatoshi {\sc Imada}
}
 
\title
{
Precise Determination of Phase Diagram for Two-Dimensional Hubbard Model with Filling- and 
Bandwidth-Control Mott Transitions: 
Grand-Canonical Path-Integral Renormalization Group Approach
}

\author
{
Shinji {\sc Watanabe}{$^1$}\footnote{
E-mail: swata@issp.u-tokyo.ac.jp} and Masatoshi {\sc Imada}{$^{1,2}$}
}

\inst
{{$^1$}
Institute for Solid State Physics, 
University of Tokyo, 5-1-5, Kashiwanoha, Kashiwa, Chiba 277-8581\\
{{$^2$}PRESTO, Japan Science and Technology Agency}
}

\recdate
{
\today
}

\abst
{ 
A new numerical algorithm for interacting fermion systems 
to treat the grand-canonical ensemble is proposed and examined by extending the path-integral renormalization group method. 
To treat the grand-canonical ensemble, 
the particle-hole transformation is applied to the Hamiltonian and basis states.
In the interaction-term projection, 
the Stratonovich-Hubbard transformation which hybridizes up and down spin
electrons is introduced. 
By using this method, the phase diagram of the two-dimensional Hubbard model 
with next-nearest-neighbor transfer is accurately 
determined 
by treating the filling-control (FC) and bandwidth-control (BC) Mott transitions on the same ground. 
A V-shaped Mott insulating phase is obtained in the plane of the chemical potential and the Coulomb interaction, 
where the transitions at the corner (BC) and the edges (FC) show contrasted characters 
with large critical fluctuations near the edges coexisting with the first-order transition at the corner. 
This contrasted behavior is shown to be consistent with the V-shape structure of the phase boundary 
because of
a general relation, in which the slope of the metal-insulator transition line in the phase diagram is expressed by thermodynamic quantities.   
The V-shaped 
opening of the Mott gap is favorably compared with the experimental results of the transition metal oxides.
}

\kword
{
path integral renormalization group method (PIRG), 
grand-canonical ensemble, GPIRG, 
geometrical frustration, 
Mott transition, 
two-dimensional Hubbard model
}

\begin{document}
\sloppy
\maketitle

\section{Introduction}
The nature of the system 
in which quantum fluctuations and electron correlations 
play essential roles is one of the main subjects  
in condensed matter physics. 
When the kinetic energy and the Coulomb repulsion 
compete severely, 
the ground state of many-body electron systems can be highly nontrivial. 
Metal-insulator transitions driven by the electron correlation 
postulated by Mott~\cite{Mott} provides a typical example of 
such a nontrivial behavior~\cite{IFT}. 

In the transition to the Mott insulator, it is known that there exist two different basic routes to control the competition of 
the interaction and kinetic energies~\cite{IFT}.  
One is the control by bandwidth (relative to the Coulomb repulsion) and the other is filling.  
Controls by these two parameters can be found in a lot of examples in real materials including transition metal compounds~\cite{IFT}, 
organic materials~\cite{MWI} and $^3$He systems~\cite{Ishida,Saunders}. 
In spite of plenty of the experimental results, phase diagrams of the Mott insulator and metals have not been fully elucidated 
in microscopic theoretical descriptions. 

The filling-control Mott transition (FCMT) was studied at zero temperature by the quantum Monte Carlo (QMC) method~\cite{IH} in the Hubbard model
on a square lattice.~\cite{FI,FI1993}  The transition shows a continuous character with a singular divergence of the compressibility and 
critical divergence of the antiferromagnetic correlation length.  
The bandwidth-control Mott transition (BCMT) was studied also at zero temperature by 
the path-integral renormalization group (PIRG) method.~\cite{MWI,KI}
In contrast to the FCMT, the BCMT shows a first-order transition, although a naive expectation is that the continuous FCMT anticipates also the continuous BCMT.  
This contrast is essentially consistent with the trend of the experimental observations cited above.
Mott has originally proposed a first-order BCMT because of the role of the long-ranged part of the Coulomb interaction~\cite{Mott}. However, the numerical result shows that the first-order transition takes place even with the onsite interaction only.  
In this circumstance, it is desired to clarify basic properties of the BCMT 
and the FCMT in a unified way to further elucidate the contrast. 

The Hubbard model has been studied for a long time as a minimal model to represent the essence of the Mott insulator and metals with 
their transitions.  
However, if the Hubbard model is defined on a bipartite lattice, the insulating gap is believed to open even at infinitesimally 
small onsite Coulomb interaction, thereby generic feature of BCMT cannot be studied in this case.

When the lattice structure of the system is ``geometrically frustrated", meaning a nonbipartite structure with the 
appearance of frustration effects in the antiferromagnetic spin exchange, the BCMT is expected to occur at a nonzero 
onsite Coulomb interaction, which enables the study of BCMT and FCMT with their comparison on the same ground.  
The Hubbard model on the square lattice with nearest-neighbor (NN)
and next-nearest-neighbor (NNN) hoppings (HMSL) 
is a typical prototype for such a system, since 
this model contains the triangular network of hopping integrals, where 
the magnetic frustration 
between the NN and NNN exchange interactions arises~\cite{anderson}. 
This model was also proposed to be a minimal model 
of the high-$T_{c}$ superconductors in a certain parameter regime near the Mott insulator. 
The LDA calculations support that the cuprates are approximately described 
by the HMSL~\cite{expparam,Andersen}. 
Among the high-$T_c$ cuprate superconductors, (La,Sr)$_2$CuO$_4$ compounds are fitted 
by the parameters of the NNN hopping in unit of the NN hopping, 
$-0.2$, 
while the YBa$_2$Cu$_3$O$_{7-\delta}$ as well as Bi compounds are fitted by 
$-0.3 \sim -0.5$.
On more general grounds, the nature of the electronic states around the Mott 
transition of the HMSL with nonzero NNN hoppings has been a intriguing issue by its own right. 
Indeed, with the frustration effects, 
the ground state can be highly nontrivial, 
since spin entropy is not easily released 
because of the competing magnetic interactions.

Numerical methods offer potential tools for studying strongly correlated systems such as the Hubbard model.
However, the phase diagram of the correlated metals and the Mott insulators in the Hubbard-type models has never been fully
clarified until recently, partially because the geometrical frustration effect causes serious difficulties in numerical approaches.  
Recently, a new numerical algorithm, 
the path-integral renormalization group (PIRG), 
has been developed, which enables us 
to obtain the ground state accurately 
even in the system with strong frustration effects~\cite{PIRG1,PIRG2}.
A remarkable point is that 
the PIRG does not suffer from the negative-sign problem 
and can be applied to any type of Hamiltonian in any dimension. 
This method improves the ground state from the Hartree-Fock state 
by increasing non-orthogonal basis functions and performing the renormalization 
in the imaginary-time direction so that 
electron-correlation effects and quantum fluctuations 
are taken into account systematically. 
To reach the exact ground state of finite systems in a controlled way, 
the zero variance limit is taken, which 
corresponds to the extrapolation of the truncated Hilbert space 
to the exact one. 

In this paper, we extend the algorithm of the PIRG to   
the grand-canonical ensemble. 
This grand-canonical 
path-integral renormalization group (GPIRG) method 
allows us to study the FCMT and BCMT within a single numerical approach in an efficient way. 
We determine the ground-state phase diagram of the HMSL 
in the plane of 
the chemical potential and bandwidth.  A V-shaped Mott insulating phase is identified.
We derive a general equation, in which the slope of the metal-insulator transition line 
in the phase diagram is expressed by the ratio of the jump of
the double occupancy to that of the electron density at the first-order transition point.  
We also derive an equation which connects the slope  to the charge compressibility 
when the continuous transition occurs. 
By analyzing these relations and the chemical-potential dependence on the electron density, 
we reach a consistent picture for the V-shaped phase boundary and a sharp contrast of the FCMT with the first-order BCMT. 
The GPIRG also allows us to 
estimate the Coulomb-interaction dependence of the charge gap
accurately.

The organization of this paper is as follows: 
In \S2, the framework of the GPIRG method is presented 
and  technical details for the implementation are described. 
In \S3, the algorithm of the GPIRG method is examined by comparing with 
other methods such as the exact diagonalization and the QMC methods. 
We show that the GPIRG is useful in calculating 
the chemical-potential dependence of physical quantities. 
The results of the HMSL by the GPIRG are reported in  \S4. 
The nature of the FCMT and the BCMT is discussed in detail. 
The summary is given in \S5. 

\section{Grand-Canonical Path Integral Renormalization Group Algorithm}

In this section, we propose the grand-canonical path-integral renormalization 
group (GPIRG) method. 
The GPIRG is useful for studying the chemical-potential dependence of 
physical quantities as described below.
Though this algorithm can be applied to a general form of electronic Hamiltonian on a lattice, 
we show the formalism for the Hubbard model as a typical example. 
First we describe the key elements of the GPIRG method from \S2.1 to \S2.4: 
We introduce the particle-hole transformation in \S2.1 and 
the path-integral operation in the Slater-determinant representation 
is explained in \S2.2. 
The truncation and optimization procedure, 
and the extrapolation procedure of physical quantities 
are explained in \S2.3 and \S2.4, respectively. 
The whole procedure of the GPIRG is summarized in \S2.5. 
Technical details for the implementation of the GPIRG are noted 
in \S2.6 and \S2.7. 

\subsection{Model and Particle-hole Transformation}

The Hubbard model which we consider is 
\begin{eqnarray}
H&=&-\sum_{{\langle i,j \rangle}\sigma}
t_{ij}
\left(
c^{\dagger}_{i\sigma}c_{j\sigma} + c^{\dagger}_{j\sigma}c_{i\sigma}
\right)
-\mu \sum_{i\sigma}n_{i\sigma}
\nonumber \\
& &+U\sum_{i}
\left( n_{i\uparrow} -\frac{1}{2}\right)
\left( n_{i\downarrow}-\frac{1}{2}\right),         
\label{eq:Hamil}
\end{eqnarray}
where $c_{i\sigma}$ $(c^{\dagger}_{i\sigma})$
is the annihilation (creation) operator on the $i$-th site 
with spin $\sigma$ and $n_{i\sigma}=c^{\dagger}_{i\sigma}c_{i\sigma}$
in the $N$-lattice system. 
The transfer integral is taken as 
$t_{ij}=t$ for the nearest-neighbor sites and 
$t_{ij}=t'$ for the next-nearest-neighbor sites. 
Throughout this paper, we take $t$ as the energy unit. 
The total electron number is 
\begin{eqnarray}
N_{e}=\sum_{i\sigma}^{N}\langle n_{i\sigma} \rangle, 
\nonumber
\end{eqnarray}
where $\langle ... \rangle$ represents the ground-state expectation value.

The canonical transformation~\cite{YokoShiba} 
is introduced as 
\begin{eqnarray}
c_{k\uparrow} \to c_{k}, 
\nonumber
\\
c_{-k\downarrow} \to d_{k}^{\dagger}.
\label{eq:PHtrans}
\end{eqnarray}

In terms of the new operators, the Hamiltonian is represented as 
\begin{eqnarray}
H&=&
H_{K}+H_{U}-\left(\frac{U}{4}+\mu\right)N,
\label{eq:hamil2}
\\
H_{K}&=&
-\sum_{\langle i,j \rangle}
t_{ij}
\left(
c_{i}^{\dagger}c_{j}+c_{j}^{\dagger}c_{i}
\right)
+\left(\frac{U}{2}-\mu\right)
\sum_{i}c_{i}^{\dagger}c_{i}, 
\nonumber \\
& &
+\sum_{\langle i,j \rangle}
t_{ij}
\left(
d_{i}^{\dagger}d_{j}+d_{j}^{\dagger}d_{i}
\right)
+\left(\frac{U}{2}+\mu\right)
\sum_{i}d_{i}^{\dagger}d_{i}, 
\nonumber \\
H_{U}
&=& -U\sum_{i}c_{i}^{\dagger}c_{i}d_{i}^{\dagger}d_{i}.
\nonumber
\end{eqnarray}

The total number of electrons is related to the difference 
of $c$ and $d$ particles as
\begin{eqnarray}
N_{e}=N+\sum_{i=1}^{N}\langle c_{i}^{\dagger}c_{i}
-d_{i}^{\dagger}d_{i}\rangle. 
\label{eq:nele}
\end{eqnarray}
The magnetization is given by 
\begin{eqnarray}
2S^{z}&=&\sum_{i=1}^{N}\langle c_{i\uparrow}^{\dagger}c_{i\uparrow}
-c_{i\downarrow}^{\dagger}c_{i\downarrow}
\rangle, 
\nonumber \\
&=& -N +M, 
\label{eq:mag}
\end{eqnarray}
where $M$ is the total number of $c$ and $d$ particles, 
\begin{eqnarray}
M=\sum_{i=1}^{N}\langle c_{i}^{\dagger}c_{i} 
+d_{i}^{\dagger}d_{i}\rangle. 
\nonumber
\end{eqnarray}
%

\subsection{Projection procedure of kinetic and interaction terms 
in the Slater determinant basis}
\subsubsection{path-integral operation}
In the GPIRG, numerical renormalization is performed in the imaginary-time
direction. 
The ground state $|\psi_g\rangle$ is obtained by 
\begin{eqnarray}
|\psi_g\rangle=\lim_{\tau \to \infty}\exp[-\tau H]|\psi_{0}\rangle, 
\label{eq:GS1}
\end{eqnarray}
where $|\psi_{0}\rangle$ is an initial state which is not orthogonal to 
$|\psi_g\rangle$. 
Practically, following Feynmann's path-integral formalism, 
the projection procedure is performed 
by taking sufficiently small $\Delta_{\tau}$:
\begin{eqnarray}
|\psi_g\rangle=\lim_{\Delta_\tau\to 0}\lim_{n\to\infty}
\left(\exp[-\Delta_{\tau} H]\right)^{n}|\psi_{0}\rangle, 
\nonumber
\end{eqnarray}
where we take $\tau \equiv \Delta_{\tau} n$ sufficiently large.

The projection operator can be divided into the 
kinetic term and the interaction term approximately:
\begin{eqnarray}
{\rm e}^{-\Delta_{\tau} \left(H_{K}+H_{U}\right)}
={\rm e}^{-\Delta_{\tau} H_{K}}{\rm e}^{-\Delta_{\tau} H_{U}}
+O\left(\Delta_{\tau}^{2}\right). 
\label{eq:ST}
\end{eqnarray}
In this paper, we assume that the basis states are the Slater determinants. 
By using eq.~(\ref{eq:ST}) and the basis set of the Slater determinants, 
the projection to the ground state corresponding to eq.~(\ref{eq:GS1}) 
is performed as described below. 

\subsubsection{Slater determinant basis}
We use the notation $|\phi\rangle$ to represent a Slater determinant. 
The Slater determinant which is a single-particle 
state is represented by the $2N \times M$ matrix, $\phi$:
\begin{eqnarray}
|\phi\rangle = {\prod^{M}_{k=1}}
\left(
\sum_{i=1}^{2N}[\phi]_{ik}\tilde{c}^{\dagger}_{i}
\right)
|0\rangle,
\end{eqnarray}
where 
$\tilde{c}_{i}=c_{i}$ for $i=1,...,N$ site and 
$\tilde{c}_{i}=d_{i-N}$ for $i=N+1,...,2N$ site.
The difference from the PIRG is that there appear the off-diagonal matrix
elements $[\phi]_{ik}$ for $i=1,...,N$ and $k=N+1,...,2N$ and 
for $i=N+1,...,2N$ and $k=1,...,N$, which hybridize $c$ and $d$ particles 
in the GPIRG. 

We note that the off-diagonal elements of $[\phi]_{ik}$
correspond to the superconducting 
order parameter, i.e., 
$\langle c_{k}^{\dagger}d_{k} \rangle=
\langle c_{k\uparrow}^{\dagger}c_{-k\downarrow}^{\dagger} \rangle$
in the ground-canonical ensemble. 
As will be explained below, the GPIRG offers a systematic improvement 
of a  mean-field solution 
such as the Hartree-Fock or the variational Monte Carlo method.
For example, 
we can set the BCS wavefunction as an initial state of the GPIRG 
and the fluctuation beyond it can be taken into account 
as the GPIRG calculation proceeds. 
In Appendix~A, 
the BCS wavefunction with various symmetries 
is given in the GPIRG framework 
and we make some comments on it.

We note that if we take the $2N \times N$ Slater matrix, 
it is nothing but to specify the subspace with $S^{z}=0$ 
from eq.~(\ref{eq:mag}).  
Hereafter we consider the $S^{z}=0$ subspace and hence 
take the $2N \times N$ Slater matrix, $[\phi]_{ik}$.
The constraint $S^{z}=0$ does not restrict our calculation and even the ferromagnetic state 
can be represented in principle in the form of the magnetization in the $xy$ plane.

\subsubsection{interaction-term projection}
The projection to the ground state is performed by using 
eq.~(\ref{eq:ST}) in the Slater-determinant basis. 
Though the interaction introduces a many-body term, the projection 
by the interaction term can be transformed into the sum of two Slater determinants by using the Stratonovich-Hubbard (SH) transformation. 
As seen in eq.~(\ref{eq:nele}), 
the total electron number is represented by the difference 
between $c$ and $d$ particle numbers. 
Then, we see that if the numbers of $c$ and $d$ particles 
change according to $\mu$, the algorithm allows the grand-canonical 
framework at a fixed $\mu$. 
Hence, we introduce the SH transformation 
which hybridizes $c$ and $d$ particles as follows: 
\begin{eqnarray}
\exp\left[-{\Delta_{\tau}}U\left\{
\frac{1}{2}\left(c^{\dagger}_{i}c_{i}+d^{\dagger}_{i}d_{i}\right)
-c^{\dagger}_{i}c_{i}d^{\dagger}_{i}d_{i}
\right\}\right]
\nonumber \\
=\frac{1}{2}\sum_{s=\pm 1} 
\exp\left[{\rm i}{\beta}s
\left(
c_{i}^{\dagger}d_{i}+d_{i}^{\dagger}c_{i}
\right)
\right],
\label{eq:SH}
\end{eqnarray}
where 
$
\beta=\cos^{-1}\left[\exp(-{\Delta_{\tau}}U/2)\right]
$ 
for ${\Delta_{\tau}}U>0$. 
Here, $s=\pm 1$ are the SH variables.
A proof of eq.~(\ref{eq:SH}) is given in Appendix~B. 
The above projection has the off-diagonal form 
with respect to $c$ and $d$ operators.

We also note that 
the SH transformation with the diagonal form 
for the attractive Hubbard model~\cite{Hirsh} 
is given as 
\begin{eqnarray}
\exp\left[\Delta_{\tau} Uc^{\dagger}_{i}c_{i}d^{\dagger}_{i}d_{i}\right]
=\frac{1}{2}\sum_{s=\pm 1}
\exp\left(-2\alpha s -\frac{\Delta_{\tau} U}{2}\right)
\nonumber \\
\times
\exp\left[\left( 
2\alpha s+\frac{\Delta_{\tau} U}{2}
\right)c^{\dagger}_{i}c_{i}\right]
\exp\left[\left( 
2\alpha s-\frac{\Delta_{\tau} U}{2}
\right)d^{\dagger}_{i}d_{i}\right], 
\nonumber \\
\label{eq:SH1}
\end{eqnarray}
where 
\begin{eqnarray}
\alpha=
\tanh^{-1}\sqrt{-\tanh\left(-\frac{\Delta_{\tau} U}{4}\right)}, 
\nonumber
\end{eqnarray}
for $\Delta_{\tau} U>0$. 

By using the SH transformation, 
the projection for the interaction term is performed as 
\begin{eqnarray}
\exp\left[\Delta_{\tau} Uc^{\dagger}_{i}c_{i}d^{\dagger}_{i}d_{i}\right]
|\phi\rangle
=\frac{1}{2}\left(
|\phi^{+}\rangle + |\phi^{-}\rangle
\right), 
\label{eq:SHeqpm}
\end{eqnarray}
where the sign in the right-hand side of eq.~(\ref{eq:SHeqpm}) 
specifies each SH variable $s=1$ or $-1$.
In this way, the projection by the local interaction 
generates a sum of two Slater determinants. 

By the interaction-term projection on the $i$-th site, 
the right-hand side of eq.~(\ref{eq:SHeqpm}) is expressed as follows:
\begin{eqnarray}
|\phi^{+}\rangle &=&
\prod^{M}_{k=1}
\left(
\sum_{j,j'=1}^{2N}
\left[I+X(+1)\right]_{jj'}[\phi]_{j'k}\tilde{c}^{\dagger}_{j}
\right)|0\rangle,
\nonumber \\
|\phi^{-}\rangle &=&
\prod^{M}_{k=1}
\left(
\sum_{j,j'=1}^{2N}
\left[I+X(-1)\right]_{jj'}[\phi]_{j'k}\tilde{c}^{\dagger}_{j}
\right)|0\rangle,
\nonumber
\end{eqnarray}
where $I$ is the $2N \times 2N$ unit matrix 
and 
$X(s)$ is the $2N \times 2N$ matrix. 
When the offdiagonal-type projection by eq.~(\ref{eq:SH}) 
is used in eq.~(\ref{eq:SHeqpm}), 
$X(s)$ is given by 
\begin{eqnarray}
\left[X(s)\right]_{ii}=\cos(\beta s)-1 \ &,&  
\ \left[X(s)\right]_{i,N+i}=i\sin(\beta s), 
\nonumber \\
\left[X(s)\right]_{N+i,i}=i\sin(\beta s) \ &,&  
\ \left[X(s)\right]_{N+i,N+i}=\cos(\beta s)-1, 
\nonumber
\end{eqnarray}
and 0 otherwise. 
Here, the local interaction projection is performed 
on the $i$-th site as in eq.~(\ref{eq:SHeqpm}).

If we use the diagonal-type projection by 
eq.~(\ref{eq:SH1}), $X(s)$ is given by 
\begin{eqnarray}
\left[X(s)\right]_{jj'}&=&
\exp\left[2\alpha s+\frac{\Delta_{\tau} U}{2}\right]-1, 
\nonumber  {\rm for} \  j=j'=i, \\ 
\left[X(s)\right]_{jj'}
&=&\exp\left[2\alpha s-\frac{\Delta_{\tau} U}{2}\right]-1,
\nonumber {\rm for} \ j=j'=N+i, 
\end{eqnarray}   
and 0 otherwise. 

\subsubsection{kinetic-term projection}
The kinetic-term projection is performed by multiplying the 
kinetic term of eq.~(\ref{eq:ST}) to the Slater determinant: 
\begin{eqnarray}
\exp[-\Delta_\tau H_{K}(\mu_{\rm pd})]|\phi\rangle=|{\phi}'\rangle. 
\label{eq:kinpro}
\end{eqnarray}
Then, the projection of the single-body operator transforms
a Slater determinant to another single Slater determinant. 
Here, $\mu_{\rm pd}$ is an artificially introduced chemical potential 
which is different from the real $\mu$. We call $\mu_{\rm pd}$, the pseudo chemical potential. 
By operating eq.~(\ref{eq:kinpro}) with several choices of $\mu_{\rm pd}$, 
the candidates for the ground state for given $\mu$ 
can be generated. 
Using
$\mu_{\rm pd}$ instead of the real $\mu$ in the projection process helps 
escaping from a local minima 
with a wrong electron number separated by a potential barrier from the 
real ground state.

\subsection{Truncation of the Hilbert space}
As seen in eq.~(\ref{eq:SHeqpm}) 
after the projection of the $N$-sites interaction, 
$\exp[-\Delta_\tau H_{U}]$, 
an original single Slater determinant expands to the sum over $2^{N}$ Slater 
determinants. Then it is necessary to truncate the expanded Hilbert space 
to the optimal state for the ground state, which is expressed as 
\begin{eqnarray}
|\psi\rangle= \sum_{a=1}^{L}w_{a}
|\phi_{a}\rangle, 
\label{eq:WaveF}
\end{eqnarray}
where $L$ is the number of the optimized Slater determinants and $w_{a}$ is 
the optimized coefficient. 
To perform the truncation and optimization of the states, 
we select the set of the Slater determinants 
with the lowest energy 
after each local-interaction projection, eq.~(\ref{eq:SHeqpm}) 
and the kinetic-term projection, eq.~(\ref{eq:kinpro}). 
This procedure is performed on the basis of 
the variational principle~\cite{PIRG1,PIRG2}.
Namely, after each projection we solve the generalized eigenvalue problem
\begin{eqnarray}
\sum_{b=1}^{L}\langle {\phi}_{a} |H|\phi_{b}\rangle w_{b}=
E\sum_{b=1}^{L}\langle {\phi}_{a}|\phi_{b}\rangle w_{b}, 
\label{eq:GEP}
\end{eqnarray}
and obtain the optimized set of $w_{b}$. 
To calculate 
$\langle {\phi}_{a} |H|\phi_{b}\rangle$, 
the Wick's theorem can be used, since the basis set is constructed 
by the sets of the single-particle Slater determinants.

\subsection{Variance extrapolation}
If we can store the dimension $L$ equal to the whole Hilbert space, 
eq.~(\ref{eq:WaveF}) gives the exact ground state. 
However, in practice the whole Hilbert space is not tractable for large system sizes.  Hence, we use the extrapolation procedure 
by the energy variance~\cite{Sorella2,PIRG1,PIRG2}. 
This is based on the fact that the deviation from the ground state energy 
is proportional to the energy variance
\begin{eqnarray}
{\delta}_E \propto {\Delta}_E,  
\label{eq:dedV}
\end{eqnarray}
when the optimized state of eq.~(\ref{eq:WaveF}) is a good approximation 
of the ground state.
Here, $\delta_E$ is given by 
\begin{eqnarray}
{\delta}_E =\langle H \rangle - \langle H \rangle_g,
\nonumber
\end{eqnarray}
where $\langle H \rangle$ and 
$\langle H \rangle_g$ denote the ground-state energy 
with restricted and whole Hilbert spaces, respectively. 
The energy variance is defined by 
\begin{eqnarray}
\Delta_E &=&
\frac{\left(\langle H^{2} \rangle - \langle H \rangle^{2}\right)}
{\langle H \rangle^{2}}. 
\nonumber
\end{eqnarray}
By using the relation of eq.~(\ref{eq:dedV}), the ground-state energy 
is obtained by the linear extrapolation to the $\Delta_E \to 0$ limit. 
The expectation value of the physical quantities such as 
equal-time correlation functions is also obtained by taking the 
linear extrapolation to the $\Delta_E \to 0$ limit. 

\subsection{Whole procedure of the GPIRG method}
Until the previous section, we explained the key elements of 
the algorithm. 
Here the whole procedure of the GPIRG is summarized. 

\vspace{8pt}
\noindent
1)
Let us start with the state, 
$|\psi\rangle= \sum_{a=1}^{L}w_{a}|\phi_{a}\rangle$. 

When $L=1$, the lowest-energy state 
$|\psi\rangle = |\phi_{1}\rangle$ 
is given by the Hartree-Fock solution. 
We can also set a variational basis state explicitly 
as an initial state of the GPIRG as shown in Appendix~A. 

First we choose a basis state $|\phi_{a}\rangle$ 
from $L$ stored basis states
$\{|\phi_{1}\rangle,  |\phi_{2}\rangle, ..., |\phi_{L}\rangle\}$, which 
will be operated by $\exp[-\Delta_\tau H]$. 

\vspace{8pt}
\noindent
2) The projection by the kinetic term is performed as 
\begin{eqnarray}
|\phi_{a}'\rangle={\rm e}^{-\Delta_{\tau} H_{K}(\mu_{\rm pd})}
|\phi_{a}\rangle, 
\label{eq:kinproj}
\end{eqnarray}
where $\mu_{\rm pd}$ is the pseudo chemical potential.
In practice, we set several $\mu_{\rm pd}$'s around $\mu$ 
and the kinetic-term projections are performed by changing $\mu_{\rm pd}$. 
After each operation of eq.~(\ref{eq:kinproj}), 
we solve the generalized eigenvalue problem in eq.~(\ref{eq:GEP}) 
for the two basis sets:
\begin{eqnarray}
|\phi_{1}\rangle,  |\phi_{2}\rangle, ..., |\phi_{a-1}\rangle, |\phi_{a}\rangle, 
|\phi_{a+1}\rangle, ..., |\phi_{L}\rangle, 
\label{eq:set1} \\
|\phi_{1}\rangle,  |\phi_{2}\rangle, ..., |\phi_{a-1}\rangle, |\phi_{a}'\rangle,
|\phi_{a+1}\rangle, ..., |\phi_{L}\rangle. 
\label{eq:set2}
\end{eqnarray}
We always calculate the expectation value of the Hamiltonian 
in eq.~(\ref{eq:GEP}) by using $\mu$, but not $\mu_{\rm pd}$. 
Then, we select either of the basis sets of eq.~(\ref{eq:set1}) or eq.~(\ref{eq:set2}), 
which gives the lower ground-state energy. 
By employing the pseudo chemical potentials, 
the change of the electron number is promoted toward the ground state for given $\mu$
after the GPIRG procedures of 
2) and 3) described below. 
The kinetic-term projection by eq.~(\ref{eq:kinproj}) 
 is done for each $|\phi_{a}\rangle$ for $a=1,\dots,L$.

\vspace{8pt}
\noindent
3) The projection by the local interaction term 
is performed as 
\begin{eqnarray}
|\phi_{a}^{+}\rangle+|\phi_{a}^{-}\rangle=
{\rm e}^{\Delta_{\tau} Uc^{\dagger}_{i}c_{i}d^{\dagger}_{i}d_{i}}
|\phi_{a}\rangle. 
\label{eq:localint}
\end{eqnarray}
To make the electron number change toward the ground state for given $\mu$, 
the off-diagonal-type projection in eq.~(\ref{eq:SH}) is employed. 
We solve the generalized eigenvalue problem in eq.~(\ref{eq:GEP}) 
for the three basis sets:
\begin{eqnarray}
|\phi_{1}\rangle,  |\phi_{2}\rangle, ..., |\phi_{a-1}\rangle, |\phi_{a}\rangle, 
|\phi_{a+1}\rangle, ..., |\phi_{L}\rangle, 
\label{eq:Uset1} \\
|\phi_{1}\rangle,  |\phi_{2}\rangle, ..., |\phi_{a-1}\rangle, |\phi_{a}^{+}\rangle,
|\phi_{a+1}\rangle, ..., |\phi_{L}\rangle, 
\label{eq:Uset2} \\
|\phi_{1}\rangle,  |\phi_{2}\rangle, ..., |\phi_{a-1}\rangle, |\phi_{a}^{-}\rangle,
|\phi_{a+1}\rangle, ..., |\phi_{L}\rangle. 
\label{eq:Uset3}
\end{eqnarray}
Then, we select the basis set among eq.~(\ref{eq:Uset1}), eq.~(\ref{eq:Uset2}), 
and eq.~(\ref{eq:Uset3}), 
which gives the lowest ground-state energy. 
Taking either of $|\phi_{a}^{+}\rangle$ or $|\phi_{a}^{-}\rangle$ leads to the 
change of the total-electron number. 
The local-interaction-term projection is performed on the $i=1,\dots,N$ site. 
This procedure is done for each $|\phi_{a}\rangle$ for $a=1,\dots,L$. 

\vspace{8pt}
\noindent
4) To increase the Hilbert space dimension, $L$, a new state is generated. 
Practically, we multiply the interaction term 
${\rm e}^{\Delta_{\tau} Uc^{\dagger}_{i}c_{i}d^{\dagger}_{i}d_{i}}$ 
to a state $|\phi_{a}\rangle$ 
and select either state of 
$|\phi_{a}^{+}\rangle$ or $|\phi_{a}^{-}\rangle$. 
Here, off-diagonal-type projection 
eq.~(\ref{eq:SH}) is employed, 
since the hybridization between $c$ and $d$ particles 
promotes the change of the electron number in the grand-canonical ensemble. 
The state $|\phi_{a}\rangle$ and the local site $i$ to which 
eq.~(\ref{eq:SH}) is operated are selected arbitrary by using random number, 
respectively. 
We also select either state, $|\phi_{a}^{+}\rangle$ or $|\phi_{a}^{-}\rangle$ 
by using the random number. 

\vspace{8pt}
\noindent 
5) The procedure from 2) to 4) is repeated until the sufficient 
number of basis $L$ is reached. 
Typically, it is sufficient to take $L=200 \sim 400$. 

\vspace{8pt}
\noindent
6) To reach the true ground state in finite systems, we extrapolate 
the physical quantiles such as the ground-state energy and 
correlation functions to the zero-variance limit as explained in \S2.4. 
Namely, in the plane of the physical quantity 
$\langle A \rangle$ and the energy variance ${\Delta}_E$, 
the linear dependence appears in the case of the well converged states.
Then, the extrapolated value by linear extrapolation is obtained 
finally.

Here we note that in the GPIRG the error bar can appear by the two procedures: 
One is the variance extrapolation in each system size. 
If the linearity in the fitting as a function of the energy variance becomes 
worse, the error bars increase. 
The other comes from  the extrapolation to the thermodynamic limit 
by assuming the finite-size corrections. 
The latter is a common origin of error bars to all the other numerical methods 
for finite-size systems, while the former is specific to this GPIRG method. 
The error bars in our analyses are given from the combination of these two types 
of error bars. 

In the interaction-term projection, 
we employ the diagonal-type projection in eq.~(\ref{eq:SH1}) 
on the sites selected by the random number, 
in addition to the off-diagonal-type projection in eq.~(\ref{eq:SH}), 
both of which with an optimized combination make the convergence faster empirically. 
This is due to the fact that 
the diagonal-type projection gives the smaller variance 
by improving the basis state 
within the fixed electron number, 
while the off-diagonal-type projection tends to change the states 
with different electron numbers, which leads to the larger variance. 

We also note that in the GPIRG the negative-sign problem 
which often becomes serious in the QMC method 
does not appear, 
since the variational wave function is explicitly given, 
$|\psi\rangle=\sum_{a=1}^{L}w_{a}|\phi_{a}\rangle$. 
Hence, the GPIRG 
can be applied to any lattice structures such as one-, two-, and 
three-dimensional systems with any boundary conditions.

\subsection{Implementation of GPIRG}
For implementation of the GPIRG procedure, 
the techniques developed in the QMC method~\cite{IH} are available. 
The matrix elements for Hamiltonian in eq.~(\ref{eq:GEP}) 
are calculated from the single-particle Green's function by using 
Wick's theorem, because the Slater determinant is a single-particle 
state. 
The inner product of two Slater determinants is given by 
\begin{eqnarray}
\langle\phi_{a}|\phi_{b}\rangle
=\det\left(^{t}[\phi_{a}][\phi_{b}]\right), 
\label{eq:det1}
\end{eqnarray}
where $[\phi_{a}]$ is the $2N \times N$ 
Slater matrix in eq.~(\ref{eq:det1}). 
The single-particle Green's function is defined by 
\begin{eqnarray}
\left[G^{ab}\right]_{ij}&=&\frac{\langle \phi_{a}|
\tilde{c}_{i}^{\dagger}\tilde{c}_{j}
|\phi_{b}\rangle}
{\langle \phi_{a}|\phi_{b}\rangle}, 
\label{eq:green1}
\end{eqnarray}
where $i$ and $j$ specify the site index from 1 to $2N$ and 
$\tilde{c}_{i}=c_{i}$ for $i=1,...,N$ site and 
$\tilde{c}_{i}=d_{i-N}$ for $i=N+1,...,2N$ site.
The $2N \times 2N$ matrix of the Green's function is calculated as 
\begin{eqnarray}
\left[G^{ab}\right]_{ij}
=\sum_{i'=1}^{2N}\sum_{j'=1}^{2N}
[\phi_{b}]_{ii'}[g^{ab}]_{i'j'}[\phi_{a}]_{j'j},
\label{eq:green2}
\end{eqnarray}
where 
\begin{eqnarray}
\left[g^{ab}\right]=\left(^{t}[\phi_{a}][\phi_{b}]\right)^{-1}. 
\nonumber
\end{eqnarray}
The physical quantities such as equal-time correlation function 
can be calculated by using the Wick's theorem and the 
Green's function, eq.~(\ref{eq:green1}). 
>From the computational point of view, 
the calculations for the determinant in eq.~(\ref{eq:det1})
and the inverse matrix in eq.~(\ref{eq:green2}) 
cost computational time proportional to $N$. 
However, this can be saved by taking the following procedures.

\subsection{Update of inner product and Green's function}
Let us consider the case that $|\phi_{b}\rangle$ is 
updated by the interaction-term projection in eq.~(\ref{eq:localint}): 
\begin{eqnarray}
\left[ 
\phi_{b}
\right]_{nk}
\to
\left[
\phi_{b}'
\right]_{nk}
=
\sum_{l=1}^{2N}\left[I+X(s)\right]_{nl}
\left[
\phi_{b}
\right]_{lk}. 
\end{eqnarray}
Hereafter we consider the case that the projection is performed on 
the $i$-th site as in eq.~(\ref{eq:localint}) 
and write $j=N+i$, 
until the end of this section \S2.7. 

The updated inner product is given by 
\begin{eqnarray}
\langle\phi_{a}|\phi'_{b}\rangle
=D\langle\phi_{a}|\phi_{b}\rangle, 
\label{eq:inpro1}
\end{eqnarray}
where $D\equiv\det(I+GX)=\det(I+XG)$ is given by 
\begin{eqnarray}
D=-F_{ij}F_{ji}+(1+F_{ii})(1+F_{jj}),
\label{eq:det}
\end{eqnarray}
with 
\begin{eqnarray}
F_{in}&=&G_{ii}X_{in}+G_{ij}X_{jn},
\nonumber \\
F_{jn}&=&G_{ji}X_{in}+G_{jj}X_{jn}.
\label{eq:F} 
\end{eqnarray}

The updated Green's function 
\begin{eqnarray}
\left[{G'}^{ab}\right]_{nm}=\frac{\langle \phi_{a}|
\tilde{c}_{n}^{\dagger}\tilde{c}_{m}
|\phi_{b}'\rangle}
{\langle \phi_{a}|\phi_{b}'\rangle},
\end{eqnarray}
is represented in the notation with $[~~^{ab}]$ omitted, as follows: 
\begin{eqnarray}
G'_{nm}=\tilde{G}_{nm}&+&
\delta_{ni}
\left(
X_{ii}\tilde{G}_{im}+X_{ij}\tilde{G}_{jm}
\right)
\nonumber \\
&+&
\delta_{nj}
\left(
X_{ji}\tilde{G}_{im}+X_{jj}\tilde{G}_{jm}
\right), 
\label{eq:updtG1}
\end{eqnarray}
where 
\begin{eqnarray}
\tilde{G}_{nm}&=&G_{nm}
-\frac{1}{D}\left[
(G_{ni}X_{ii}+G_{nj}X_{ji})
\right.
\nonumber \\
&\times&
\left.
\left\{1+F_{jj})G_{im}-(-1)^{i+j}F_{ij}G_{jm}\right\}
\right.
 \nonumber \\
&+&
\left.
(G_{ni}X_{ij}+G_{nj}X_{jj})
\right.
\nonumber \\
&\times&
\left.
\left\{(1+F_{ii})G_{jm}-(-1)^{i+j}F_{ji}G_{im}\right\}
\right]. 
\label{eq:tG}
\end{eqnarray}
By using eqs.~(\ref{eq:inpro1}), ~(\ref{eq:updtG1}) and ~(\ref{eq:tG}), 
the updated inner product and Green's function $G'$ can be obtained 
by using $G$. 
Namely, if the Green's function $G$ before updating is known, 
the updated inner product and Green's function $G'$ can be obtained by 
eq.~(\ref{eq:inpro1}) and eq.~(\ref{eq:updtG1}), respectively 
without calculating the determinant in eq.~(\ref{eq:det1}) and 
the inverse matrix in eq.~(\ref{eq:green2}).

In the case that $\langle \phi_{a}|$ is updated 
in eq.~(\ref{eq:localint}), 
\begin{eqnarray}
\left[ 
\phi_{a}
\right]_{ij}
\to
\left[
\phi_{a}'
\right]_{ij}
=
\sum_{n=1}^{2N}\left[I+X(s)\right]_{in}
\left[
\phi_{a}
\right]_{nj}, 
\nonumber
\end{eqnarray}
the updated inner product 
$\langle\phi'_{a}|\phi_{b}\rangle$ 
is obtained in eqs.~(\ref{eq:inpro1}), 
(\ref{eq:det}) and (\ref{eq:F}) 
with $X$ replaced by $^{t}X$. 
The updated Green's function 
\begin{eqnarray}
\left[{G'}^{ab}\right]_{nm}=\frac{\langle \phi_{a}'|
\tilde{c}_{n}^{\dagger}\tilde{c}_{m}
|\phi_{b}\rangle}
{\langle \phi_{a}'|\phi_{b}\rangle}
\end{eqnarray}
is obtained in the form with $[~~^{ab}]$ omitted 
as follows:
\begin{eqnarray}
G_{nm}'= \tilde{G}_{nm}
&+& 
\left(
\tilde{G}_{ni}X_{ii}+\tilde{G}_{nj}X_{ij}
\right)\delta_{in}
\nonumber \\
&+& 
\left(
\tilde{G}_{ni}X_{ji}+\tilde{G}_{nj}X_{jj}
\right)\delta_{jm}, 
\label{eq:updtG2}
\end{eqnarray}
where $\tilde{G}_{nm}$ is calculated in eqs.~(\ref{eq:det}), (\ref{eq:F})
and (\ref{eq:tG}) with $X$ replaced by $^{t}X$. 

By using eqs.~(\ref{eq:updtG1}) and (\ref{eq:updtG2}), we can 
calculate the matrix element 
$
\langle {\phi}_{a} |H|\phi_{b}\rangle
$
and 
$
\langle {\phi}_{a}|\phi_{b}\rangle 
$
from the Green's function of the previous step 
after each local-interaction-term projection, eq.~(\ref{eq:localint}). 

We note that these expressions for update of the inner product 
and the Green's function can be used in the general case where 
the interaction term affects the $i$ and $j$ sites. 
Namely, these are not restricted to the GPIRG method, but can be used 
in the PIRG and QMC methods: 
For example, after multiplication of 
the Slater determinants and the interaction term such as 
the exchange interaction and pair-hopping term 
in the multi-orbital system, eqs.~(\ref{eq:inpro1}), 
(\ref{eq:updtG1}) and (\ref{eq:updtG2}) are available for update.


\section{Examination of the GPIRG method}
The accuracy of the GPIRG method is examined by 
comparing with results by other methods 
such as the exact diagonalization, the QMC, and the PIRG methods.

\subsection{Comparison of energy}
\begin{figure}
\begin{center}
\epsfxsize=8cm \epsfbox{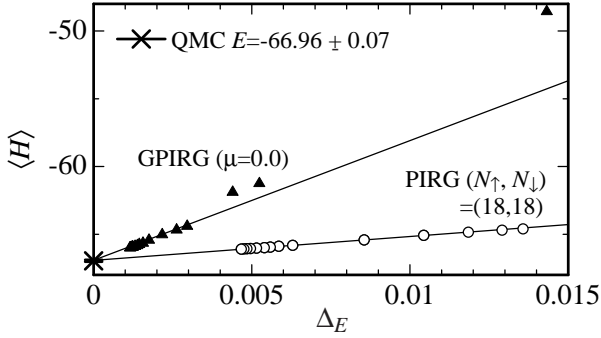}
\end{center}
\caption{
Extrapolation of the ground-state energy 
by the GPIRG (solid triangle) with $\mu=0$ and 
to the zero-variance limit 
in the HMSL for $t=1.0$, $U=4.0$ on the $N=6{\times}6$ lattice 
under the periodic boundary condition. 
The PIRG data for $(N_\uparrow,N_\downarrow)=(18,18)$ 
is represented by open circles. 
The QMC result is shown by cross on the $\langle H\rangle$ axis.
}
\label{QMCPIRG3}
\end{figure}

\begin{figure}
\begin{center}
\epsfxsize=7.8cm \epsfbox{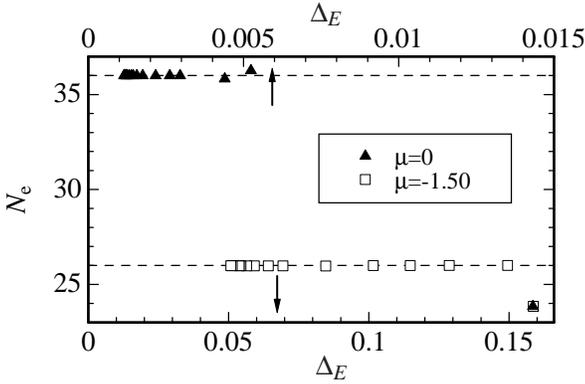} 
\end{center}
\caption{
Extrapolation of the total electron number $N_{e}$ 
by the GPIRG for $\mu=-1.50$ (open squares) and 
for $\mu=0.0$ (solid triangles) 
in the HMSL for $t=1.0$, $U=4.0$ on the $N=6{\times}6$ lattice 
under the periodic boundary condition. 
}
\label{QMCPIRG2}
\end{figure}

First we compare the ground-state energy between the GPIRG and other methods. 
In Fig.~\ref{QMCPIRG3} the extrapolation of the ground-state energy 
by the GPIRG with $\mu=0.0$ 
for $t=1.0$, $t'=0.0$ and $U=4.0$ on the $6 \times 6$ HMSL 
under the periodic boundary condition 
is shown by filled triangles. 
The total electron number 
$N_{e}$ versus(vs.) variance for the corresponding data is 
shown in Fig.~\ref{QMCPIRG2} by filled triangles, and 
we see that this state is converged into $N_{e}=36$ 
after the variance extrapolation.  
Since we consider the $S_z=0$ subspace in the GPIRG in this paper, 
it turns out that the state has the electron number 
$(N_{\uparrow},N_{\downarrow})=(18,18)$. 
The ground-state energy after variance extrapolation is 
estimated at $E=-66.94 \pm 0.05$. 
For comparison, we show the PIRG results (open circle) by setting 
$(N_{\uparrow},N_{\downarrow})=(18,18)$ in Fig.~\ref{QMCPIRG3}. 
The ground-state energy by the PIRG after variance extrapolation is 
estimated at $E=-66.92 \pm 0.04$~\cite{PIRG2}. 
A more accurate estimate indicated $E=-66.88$ with the error less than 0.01~\cite{MI}.
The number of states up to $L=320$ are shown 
in both the GPIRG and the PIRG methods. 
The ground-state energies by the GPIRG and PIRG methods are consistent 
within the error bars by the variance extrapolation. 
The QMC data, $E=-66.96 \pm 0.07$~\cite{PIRG2} 
is also plotted as the cross. 
We see that the GPIRG data and the PIRG data are consistent with 
the QMC data.

%
%
%
%

We have also checked the GPIRG with other methods 
in larger system sizes. 
For example, 
when the ground-state energy for the zero-variance limit 
of the GPIRG with $\mu=0.0$ 
is written by $E$ 
and that of the PIRG for 
$(N_{\uparrow},N_{\downarrow})=(N/2,N/2)$ 
is written by $E'$, 
the relative error defined by $(E-E')/E$ is estimated at 
$2.26 \times 10^{-4}$ for the $N=8 \times 8$ system and 
$1.98 \times 10^{-4}$ for the $N=10 \times 10$ system 
for $t=1.0$, $t'=0.0$ and $U=4.0$. 
Since it has been confirmed in ref.~\cite{PIRG2}
that the PIRG data agree with 
the QMC data for $N=8 \times 8$ and $10 \times 10$ systems
within the relative error less than $0.3 \%$, 
these results indicate that the GPIRG also provides reliable data 
comparable to these methods even in large-sized systems.

The comparison in the energy between the GPIRG and the other methods are 
summarized in Table.~\ref{comparison1}. 
We see that the relative error is less than $0.3 \%$. 

\begin{fulltable}
\begin{fulltabular}{cccc}
\hline
system & GPIRG & exact diagonalization and Monte Carlo results & relative error \\
      &  $E=\langle H \rangle$ 
& $E'=\langle H \rangle$ & $|(E-E')/E|$ \\ \hline
$6\times 2$, $5\uparrow, 5\downarrow$  & -25.70 $\pm$ 0.02 & -25.6952 & 0.00022 \\
$6\times 6$, $13\uparrow, 13\downarrow$ & -58.45 $\pm$ 0.09 &  -58.32 $\pm$ 0.072 & 0.0022 \\
$6\times 6$, $18\uparrow, 18\downarrow$  & -66.94 $\pm$ 0.05 & -66.96$\pm$0.07 & 0.00031 \\
\hline
\end{fulltabular}
\caption{
Comparison between GPIRG and exact diagonalization and Monte Carlo results 
in the ground-state energy for $t=1.0$ and $U=4.0$ 
for various-size systems 
under the periodic boundary conditions in $x$ and $y$ directions.
}
\label{comparison1}
\end{fulltable}

\subsection{Comparison of other physical quantities}
Next we compare the GPIRG method with the exact diagonalization method 
for correlation functions. 

\begin{figure}
\begin{center}
\epsfxsize=7.8cm \epsfbox{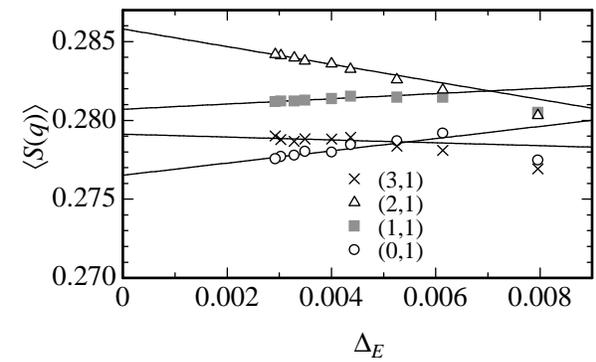} 
\end{center}
\caption{
Extrapolation of the equal-time spin correlations $S(q_x,q_y)$ 
to zero energy variance of the GPIRG method with $\mu=-1.25$ 
in the HMSL for $t=1.0$, $U=4.0$ on the $N=6{\times}2$ lattice 
under the periodic boundary condition. 
}
\label{fig:Sq1}
\end{figure}

\begin{figure}
\begin{center}
\epsfxsize=7.8cm \epsfbox{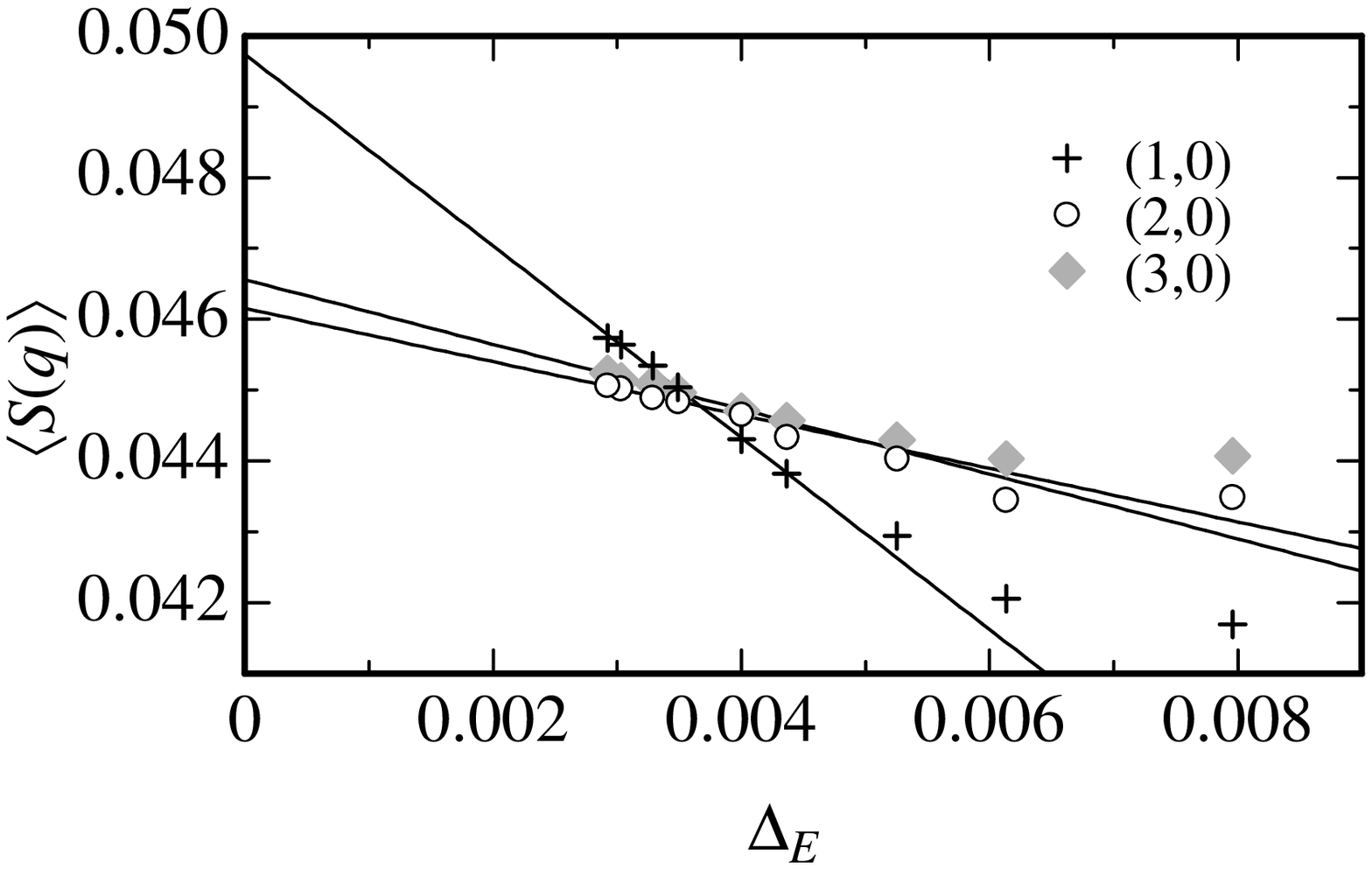} 
\end{center}
\caption{
Extrapolation of the equal-time spin correlations $S(q_x,q_y)$ 
to zero energy variance of the GPIRG method with $\mu=-1.25$ 
in the HMSL for $t=1.0$, $U=4.0$ on the $N=6{\times}2$ lattice 
under the periodic boundary condition. 
}
\label{fig:Sq2}
\end{figure}

The equal-time spin correlations in the momentum space is calculated 
from
\begin{eqnarray}
S({\bf q})=\frac{1}{3N}\sum_{i.j}^{N} 
\left\langle
{\bf S}_{i} \cdot {\bf S}_{j}
\right\rangle
{\rm e}^{{\rm i}{\bf q}\cdot{({\bf R}_{i}-{\bf R}_{j})}}, 
\label{eq:spin}
\end{eqnarray}
where ${\bf S}_{i}$ is the spin operator of the $i$-th site 
and each element of the spin is calculated from
\begin{eqnarray}
S_{i}^{x}&=&\frac{1}{2}\left( 
S_{i}^{+}+S_{i}^{-}
\right)
=\frac{1}{2}\left(
c_{i\uparrow}^{\dagger}c_{i\downarrow}+
c_{i\downarrow}^{\dagger}c_{i\uparrow}
\right), 
\nonumber \\
S_{i}^{y}&=&\frac{1}{2i}\left( 
S_{i}^{+}-S_{i}^{-}
\right) 
=\frac{1}{2}\left(
c_{i\uparrow}^{\dagger}c_{i\downarrow}-
c_{i\downarrow}^{\dagger}c_{i\uparrow}
\right), 
\nonumber \\
S_{i}^{z}&=&\frac{1}{2}\left(
n_{i\uparrow}-n_{i\downarrow}
\right). 
\nonumber 
\end{eqnarray}
The expectation value of eq.~(\ref{eq:spin}) 
is calculated by using the Wick's theorem and the Green's function, 
eq.~(\ref{eq:green1}). 
Namely, after the particle-hole transformation, eq.~(\ref{eq:PHtrans}), 
the expectation value of operators is decomposed by the Wick's theorem. 
The difference from the PIRG case is that the off-diagonal term 
for $\uparrow$ and $\downarrow$ operators 
becomes finite in the GPIRG framework; 
for example, in the representation after the particle-hole transformation, 
$\langle c_{i}d^{\dagger}_{i} \rangle
\langle d_{i}c^{\dagger}_{i} \rangle$ 
can have a nonzero value. 
We performed the GPIRG with $\mu=-1.25$ for $t=1.0$, $t'=0.0$ and $U=4.0$ 
on the $6 \times 2$ HMSL under the periodic boundary condition. 
All the extrapolations of the equal-time spin correlations 
are shown in Figs.~\ref{fig:Sq1} and~\ref{fig:Sq2}. 
The results after the extrapolation to the zero-variance limit is 
shown in Table.~\ref{Sqtable}. 
By a proper choice of the chemical potential, 
the GPIRG result indicates the particle number converges to $N_{e}=10$ which is nothing but 
$(N_{\uparrow},N_{\downarrow})=(5,5)$ electrons. 
The result of the exact diagonalization 
for $(N_{\uparrow},N_{\downarrow})=(5,5)$ electrons is shown 
for comparison in Table.~\ref{Sqtable}.  

%
%

%
%

\begin{table}
\begin{tabular}{cccc}
\hline
$(q_x,q_y)$ & GPIRG & exact diagonalization & relative error \\
\hline
(1,0) & 0.0497  & 0.0488 & 0.018  \\
(2,0) & 0.0462  & 0.0458 & 0.009  \\
(3,0) & 0.0466  & 0.0457 & 0.019  \\
(0,1) & 0.2765  & 0.277  & 0.002  \\
(1,1) & 0.2807  & 0.281  & 0.001  \\
(2,1) & 0.2858  & 0.286  & 0.001  \\
(3,1) & 0.2791  & 0.279  & 0.000  \\
\hline
\end{tabular}
\caption{
Comparison between GPIRG and exact diagonalization (ED) in the 
momentum distribution $S(q_x,q_y)$
for $t=1.0$ and $U=4.0$ 
on the $N=6 \times 2$ lattice 
for $(N_{\uparrow},N_{\downarrow})=(5,5)$ electrons 
under the periodic boundary condition.
}
\label{Sqtable}
\end{table}

\begin{figure}
\begin{center}
\epsfxsize=7.8cm \epsfbox{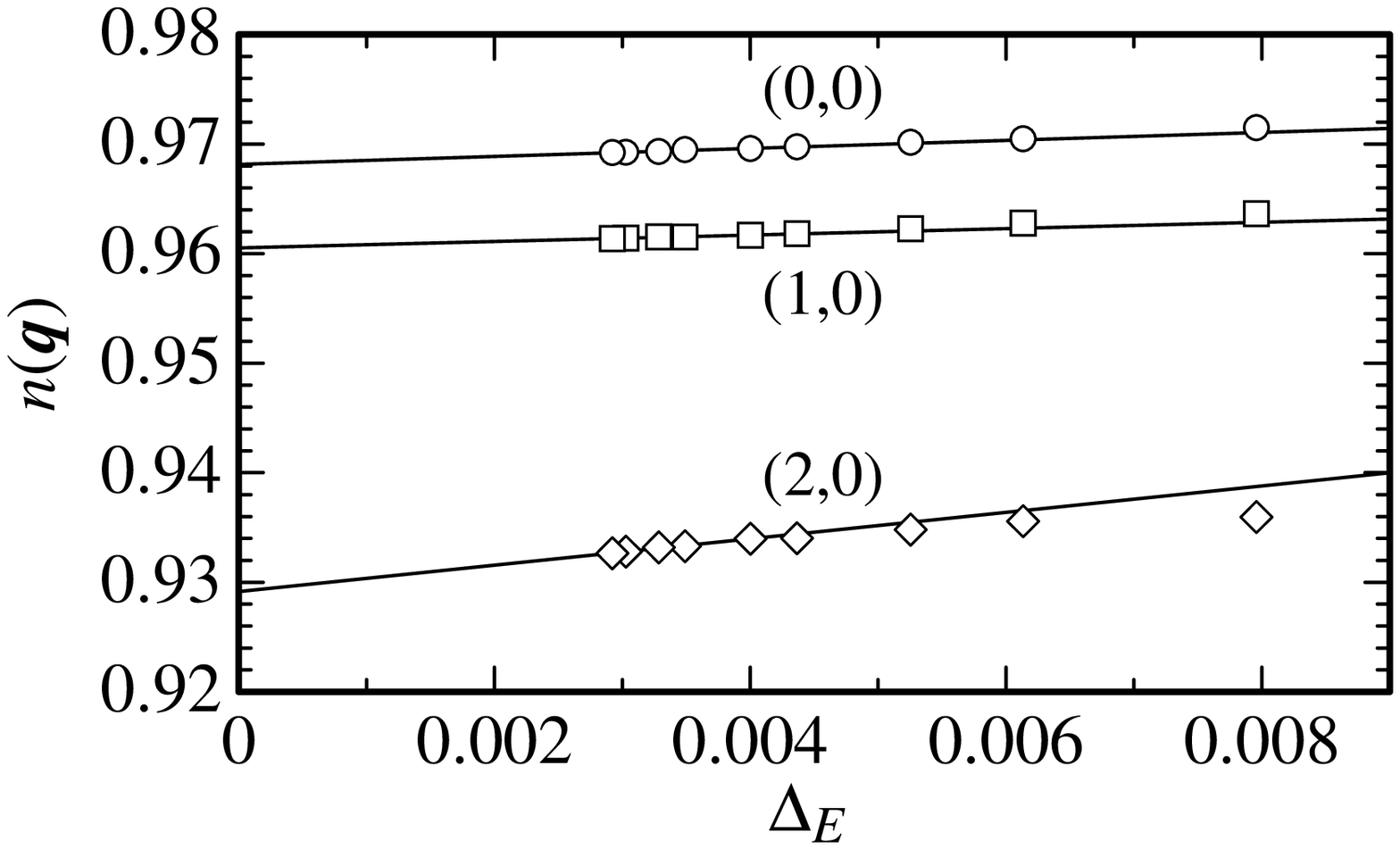} 
\end{center}
\caption{
Extrapolation of the momentum distribution functions $n(q_x,q_y)$ 
to zero energy variance of the GPIRG method with $\mu=-1.25$ 
in the HMSL for $t=1.0$, $U=4.0$ on the $N=6{\times}2$ lattice 
under the periodic boundary condition. 
}
\label{fig:mom1}
\end{figure}

\begin{figure}
\begin{center}
\epsfxsize=7.8cm \epsfbox{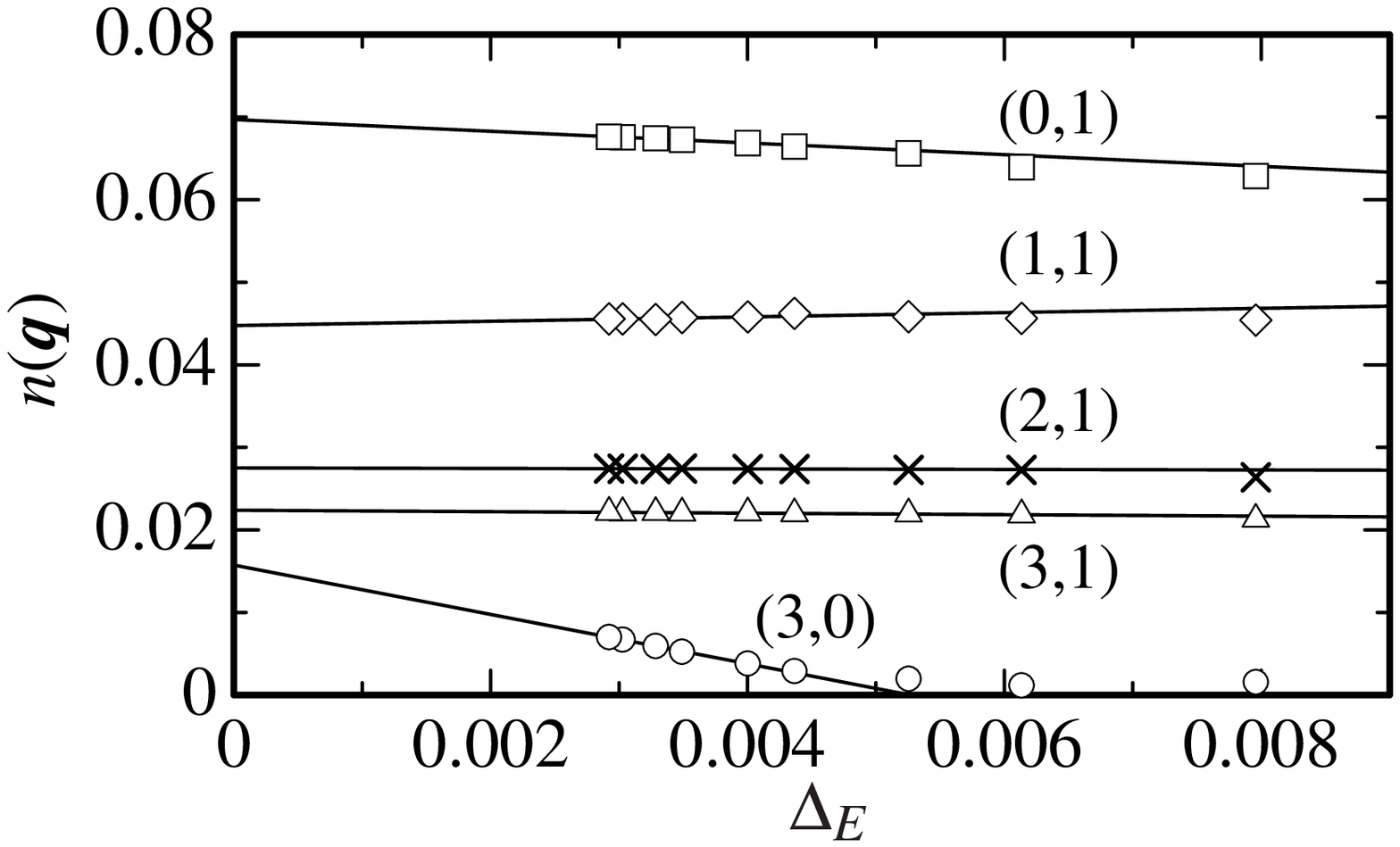} 
\end{center}
\caption{
Extrapolation of the momentum distribution functions $n(q_x,q_y)$ 
to zero energy variance of the GPIRG method with $\mu=-1.25$ 
in the HMSL for $t=1.0$, $U=4.0$ on the $N=6{\times}2$ lattice 
under the periodic boundary condition. 
}
\label{fig:mom2}
\end{figure}

The momentum-distribution function is defined by 
\begin{eqnarray}
n({\bf q})=\frac{1}{2N}\sum_{i.j}^{N} 
\left\langle
c^{\dagger}_{i\uparrow}c_{j\uparrow}
+c^{\dagger}_{j\downarrow}c_{i\downarrow}
\right\rangle
{\rm e}^{{\rm i}{\bf q}\cdot{({\bf R}_{i}-{\bf R}_{j})}}, 
\label{eq:mom}
\end{eqnarray}
where ${\bf R}_{i}$ is the vector representing the place of the 
$i$-th site.
The expectation value of eq.~(\ref{eq:mom}) 
is calculated by using the Green's function, 
eq.~(\ref{eq:green1}). 
A relatively large error is seen for $n({\bf q})$ at $(q_{x},q_{y})=(3,0)$. 
This is related to the fact that the amplitude itself is already small 
$(=0.0185)$. 
All the extrapolations of the momentum-distribution functions 
for the same system as in Fig.~\ref{fig:Sq1} 
are shown in Figs.~\ref{fig:mom1} and~\ref{fig:mom2}. 
The results after the extrapolation to the zero-variance limit 
by the GPIRG method are shown in Table.~\ref{nqtable}. 

%
%

%
%

\begin{table}
\begin{tabular}{cccc}
\hline
$(q_x,q_y)$ & GPIRG & exact diagonalization & relative error \\
\hline
(0,0) & 0.9681  & 0.9681 & 0.0000    \\
(1,0) & 0.9606  & 0.9601 & 0.0005    \\
(2,0) & 0.9291  & 0.9281 & 0.0011    \\
(3,0) & 0.0157  & 0.0185 & 0.1783    \\
(0,1) & 0.0697  & 0.06806 & 0.0235  \\
(1,1) & 0.0448  & 0.04513 & 0.0074  \\
(2,1) & 0.0275  & 0.02787 & 0.0135  \\
(3,1) & 0.0223  & 0.02281 & 0.0229  \\
\hline
\end{tabular}
\caption{
Comparison between GPIRG and exact diagonalization (ED) in the 
momentum distribution $n(q_x,q_y)$ 
for $t=1.0$ and $U=4.0$ 
on the $N=6 \times 2$ lattice 
for $(N_{\uparrow},N_{\downarrow})=(5,5)$ electrons 
under the periodic boundary condition.
}
\label{nqtable}
\end{table}

In all the equal-time spin correlations and the momentum distributions, 
except $n(3,0)$, the relative errors 
are less than $3\%$. 

\subsection{Change of total-electron number}
\begin{figure}
\begin{center}
\epsfxsize=7.8cm \epsfbox{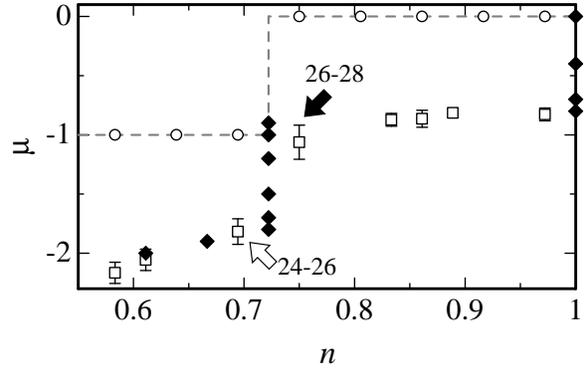}
\end{center}
\caption{
Chemical potential $\mu$ versus filling $n$ calculated by the 
QMC~\cite{FI} (open squares) and the GPIRG (solid diamonds) 
in the HMSL for $t=1.0$, $U=4.0$ on the $N=6{\times}6$ lattice 
under the periodic boundary condition. 
The open circles with dashed line is for $U=0.0$. 
The filled arrow represents $\mu$ 
calculated by QMC~\cite{FI} for canonical ensemble at 
$N_{e}=26$ and $N_{e}=28$
and open arrow represents $\mu$ calculated from $N_{e}=24$ and 
$N_{e}=26$ (see text). 
}
\label{QMCPIRG1}
\end{figure}

Next, to clarify how the grand canonical ensemble works in our GPIRG algorithm, 
we examine the relation between the chemical potential and 
the total electron number. 
In Fig.~\ref{QMCPIRG1},  
the chemical potential $\mu$ vs. filling $n=N_{e}/N$ 
by the QMC data~\cite{FI} are shown 
as open squares 
for  $t=1.0$, $t'=0.0$ and $U=4.0$ on the $N=6 \times 6$ lattice. 
Here, the chemical potential $\mu$ is calculated by 
\begin{eqnarray}
\mu(\tilde{n}) = \frac{E(N_{e})-E(N'_{e})}{N_{e}-N'_{e}}, 
\label{eq:chem}
\end{eqnarray}
where $E(N_{e})$ is the ground-state energy and 
$\tilde{n}=(N_{e}+N'_{e})/(2N)$. 
The open circles with dashed line represent $\mu$ vs. $n$ for $U=0.0$. 
The step of the dashed line appears at the closed shell, 
$N_{e}=26$ in the non-interacting case. 
The open square pointed by the 
filled arrow is the chemical potential 
calculated by using eq.~(\ref{eq:chem}) 
from the QMC data for the canonical ensemble~\cite{FI}
at $N_{e}=26$ and $N'_{e}=28$
and the open square pointed by the open arrow is obtained from the QMC data 
at $N_{e}=24$ and $N_{e}=26$. 
In the GPIRG, 
the total electron number $N_{e}$ is obtained 
as the output after the zero-variance extrapolation for given $\mu$.  
To check whether the correct $N_{e}$ is obtained for the given $\mu$ 
in the GPIRG, we start from a state at $L=1$ with $N_{e} \sim 24$ 
as an arbitrary initial state.
If the input parameter 
$\mu$ is located between filled and open arrows in 
Fig.~\ref{QMCPIRG1}, to be consistent with the QMC data, 
the converged state by the GPIRG should have 
$N_{e}=26$. 
In Fig.~\ref{QMCPIRG2}, the GPIRG results are 
shown by the open squares for 
$\mu=-1.50$ and we see that the GPIRG data are actually converged into 
the ground state with $N_{e}=26$. 
Next we set $\mu=0.0$, which is larger than the $\mu$ indicated by the filled arrow 
in Fig.~\ref{QMCPIRG1} 
and performed the GPIRG starting from the same $L=1$ state 
as the above case at $\mu=-1.50$. 
The results are shown by the filled triangles in Fig.~\ref{QMCPIRG2}, where
 the converged state has $N_{e}=36$. 
This also reproduces the correct result, since at $\mu=0.0$,  
 half filling should be realized. 
Namely, in the present $N=6 \times 6$ system for $t'=0$, 
half filling, $N_{e}=36$ 
should be realized for $\mu=0.0$.

%
%

\begin{figure}
\begin{center}
\epsfxsize=7.8cm \epsfbox{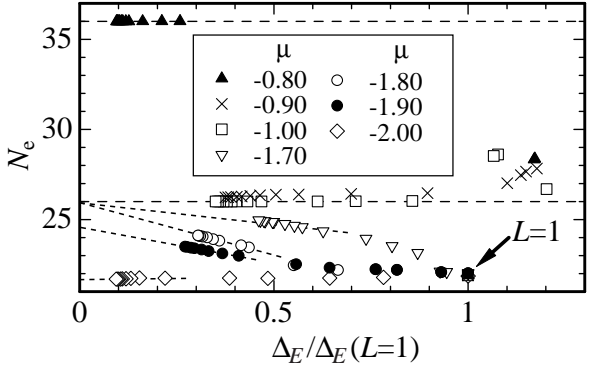} 
\end{center}
\caption{
Extrapolation of the total electron number $N_{e}$ 
by the GPIRG for $\mu=-0.8$ (solid triangle) 
$-0.90$ (cross), $-1.0$ (open square), 
$-1.70$ (open triangle), $-1.8$ (open circle),
$-1.9$ (solid circle), and 
$-2.0$ (open diamond) 
in the HMSL for $t=1.0$, $U=4.0$ on the $N=6{\times}6$ lattice 
under the periodic boundary condition. 
}
\label{QMCPIRG4}
\end{figure}

In Fig.~\ref{QMCPIRG4}, 
we show $N_{e}$ vs. the variance normalized by $\Delta_E$ at $L=1$ 
for various input parameters, $\mu$. 
Here, the same $L=1$ state is used for each simulation. 
We see that at $\mu=-0.8$, $N_{e}$ is converged to $N_{e}=36$, 
while at $\mu=-0.9$, it converges to $N_{e}=26$. 
The closed-shell electron number $N_{e}=26$ is 
stable until $\mu=-1.80$.   
While at $\mu=-1.9$ it seems to converge to 
$N_{e}=24$. 

The chemical potential vs. filling obtained 
from Fig.~\ref{QMCPIRG2} and 
Fig.~\ref{QMCPIRG4} are plotted 
by solid diamonds in Fig.~\ref{QMCPIRG1}. 
We see that the GPIRG results are consistent with the QMC results. 
As seen from Fig.~\ref{QMCPIRG1}, the electron number sensitively and correctly converges with the resolution of 
0.1 for the input parameter 
$\mu$ in the GPIRG method.
As the system size increases, 
the energy gaps of the shell structure in finite-size systems of course  
become smaller. 
This makes the potential barrier small and 
the electron number tends to change more continuously in the GPIRG calculation. 
On the other hand, in the canonical framework 
such as the PIRG~\cite{PIRG1,PIRG2}, 
the chemical potential is calculated from the subtraction of 
the ground-state energies 
with different electron numbers in eq.~(\ref{eq:chem}). 
This gives larger error in the larger system sizes. 
Hence, the GPIRG is useful to calculate the chemical-potential dependence 
of the physical quantities such as the charge gap and the $\mu$-$U$ phase 
diagram which will be shown in \S4. 

%
%


\section{Nature of Filling-Control and Bandwidth-Control 
Mott Transitions}

By using the GPIRG method 
we study the nature of the FCMT and the BCMT on the HMSL. 
We focus on $t'/t=-0.2$ where the first-order BCMT occurs~\cite{KI}, 
though the GPIRG can be applied to the larger-frustration regime. 
The GPIRG calculation is performed up to $N=10 \times 10$ lattices 
under the fully  periodic boundary condition.

\subsection{Ground-state phase diagram in plane of 
chemical potential and band width}

\begin{figure}
\begin{center}
\epsfxsize=8cm \epsfbox{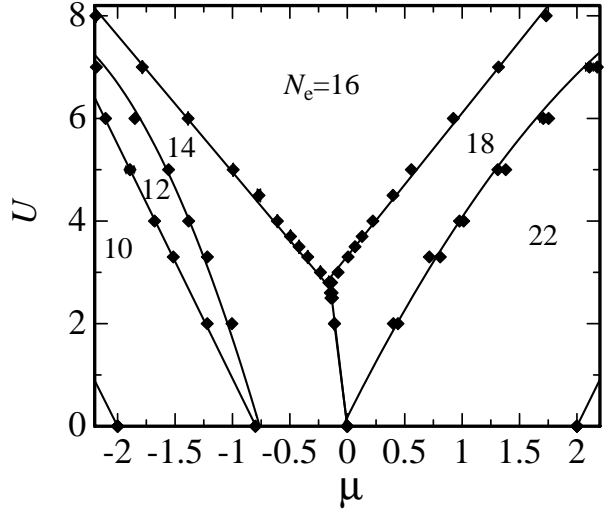}
\end{center}
\caption{
Ground-state phase diagram in the plane of 
$\mu$ and $U$ for $t=1.0$ and $t'=-0.2$ 
on the the $N=4 \times 4$ lattice. 
The solid lines are the least-square fit of each boundary 
between states with different electron number. 
}
\label{fig:mu-UN44}
\end{figure}

By performing the GPIRG with the control 
parameters $\mu$ and $U$, 
we obtain the total electron number $N_{e}$. 
Then we can construct the ground-state phase diagram 
by plotting the boundary between the states with different electron numbers 
in the plane of $\mu$ and $U$. 
Figure~\ref{fig:mu-UN44} shows the ground-state phase diagram 
for $t=1.0$ and $t'=-0.2$ on the $N=4\times4$ lattice. 
For $U=0$ the closed shells are realized at $N_{e}=10$, 14 and 22. 
The degeneracy of $\mu$ is lifted by switching on $U$. 
The solid line is the least-square fit of each boundary. 
Each area between the boundary lines corresponds to the state with each $N_{e}$. 
The central triangle area is $N_{e}=16$, i.e., half filling. 
For $U\gsim5$ the boundaries between $N_{e}=14$ and 16, and 
$N_{e}=16$ and 18, 
are remarkably linear with $U$.  This is  expected  in the large~$U$ regime, but it holds even close to the critical value $U \sim 2.8$. 
The linear opening of the gap basically comes from 
the energy cost to add an electron (a hole) 
to half filling. 
The linear-fitting lines of the boundary between $N_{e}=14$ and 16 and the boundary 
between $N_{e}=16$ and 18 for $U=2.8$, 3.0, 3.3, 3.5, 3.7, 4.0, 4.5, 5.0, 6.0, 7.0 
and 8.0 
crosses at $(\mu,U)=(-0.19,2.80)$. 

%
%

%
\begin{figure}
\begin{center}
\epsfxsize=8cm \epsfbox{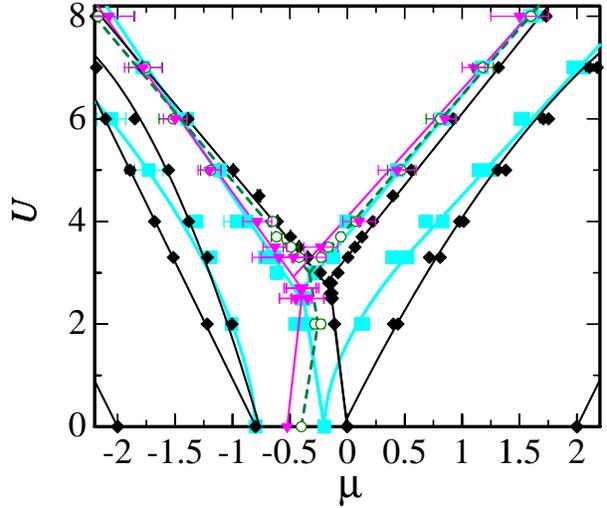} 
\end{center}
\caption{
Ground-state phase diagram in the plane of 
$\mu$ and $U$ for $t=1.0$ and $t'=-0.2$ on the square lattice. 
The data represents the boundary between different electron numbers 
in the $N=4\times 4$ (black diamond), 
$6\times 6$ (blue square), $8\times 8$ (green circle) and 
$10\times 10$ (pink triangle)
systems. 
Each line represents the least-square fit for each boundary
in the $N=4\times 4$ (black line), 
$6\times 6$ (blue line), $8\times 8$ (green dashed line) and 
$10\times 10$ (pink line) systems. 
}
\label{fig:mu-U}
\end{figure}
%

%
\begin{figure}
\begin{center}
\epsfxsize=8cm \epsfbox{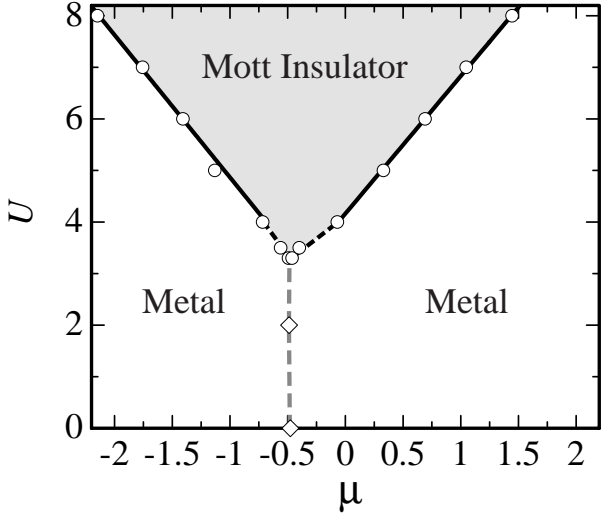}
\end{center}
\caption{
Ground-state phase diagram in the plane of 
$\mu$ and $U$ for $t=1.0$ and $t'=-0.2$ on the square lattice 
in the thermodynamic limit. 
The solid lines represent the least-square fit of metal-insulator 
transition for $U=4.0, 5.0, 6.0, 7.0$ and $8.0$. 
Gray area represents the Mott-insulator phase 
in the thermodynamic limit. 
The open diamonds represent chemical potentials for half filling
for $U=0.0$ and $U=2.0$ in the thermodynamic limit. 
By connecting those points to the BCMT point, the gray-dashed line
represents the half-filling density in the metallic phase.
The errors produced by the system-size extrapolation are 
not shown (see text). 
}
\label{fig:mu-Ubulk}
\end{figure}
%
By performing the same procedure in larger-sized systems, 
we construct 
the ground-state phase diagram in the plane of $\mu$ and $U$ 
in Fig.~\ref{fig:mu-U}. 
Here, each boundary between different 
$N_{e}$ states around $n=1$ is shown 
for the system size of $N=4 \times 4$ 
(black diamond)
as well as for $N=6 \times 6$  
(blue square).
For $N=8 \times 8$ 
(green circle)
and $N=10 \times 10$ 
(pink triangle), 
the boundary between $n=1$ and $n \ne 1$ closest to $n=1$ is shown. 
The least square fit of each data is also plotted 
for the systems with 
$N=4 \times 4$ 
(solid line), 
$6 \times 6$ 
(blue line), 
$8 \times 8$ 
(green dashed line), 
and $10 \times 10$ 
(pink line). 
To estimate the boundary of the insulator phase for $N \to \infty$, 
we extrapolate the chemical potentials between the $n=1$ state 
and the $n \ne 1$ state closest to $n=1$ 
by using the fitting function 
\begin{eqnarray}
\mu(N)=\mu + \frac{\mu'}{\sqrt{N}}.
\label{eq:mueq}
\end{eqnarray}
%
The particle number closest to half filling realized in Fig.10 is 
$N_e=32$, 62 and 98 for $6\times6$, $8\times8$ and $10\times10$ lattices, 
respectively for the hole doping. 
For the electron doping, $N_e=38$, 66 and 102 in the same order as above. 
We employ the above finite-size scaling form because it well fits the data and partly 
because the Hartree-Fock gap equation also follows this form~\cite{FI}.
The least square fit by the form eq.~(\ref{eq:mueq}) of $\mu$ for $N \to \infty$ 
at $U=4.0, 5.0, 6.0, 7.0$ and $8.0$
are shown by open circles 
in Fig.~\ref{fig:mu-Ubulk}. 

The black lines in Fig.~\ref{fig:mu-Ubulk} represent a linear fitting of these open circles,
which well represent the metal-insulator boundaries in the thermodynamic limit 
and two boundaries meet at $(\mu, U)=(-0.38,3.15)$. 
The gray area inside the thick solid lines is 
the Mott insulator phase. 
Although the error bars arising from the system-size 
extrapolation become relatively large for small $U$ $(3.3 \lsim U < 4)$ regime 
(see also Fig.~\ref{fig:mu-U}), they are typically less than 0.1.
In Fig.~\ref{fig:mu-Ubulk} 
we omit plotting the error bars arising from the system-size 
extrapolation. 
The open diamonds represent chemical potentials for half filling
for $U=0.0$ and $U=2.0$ in the thermodynamic limit. 
The gray-dashed line connects those points to 
the bottom of the Mott insulator phase, namely, the BCMT point. 
We see that the Mott insulator phase closes at 
$U_{c}=3.3$ 
in Fig.~\ref{fig:mu-Ubulk}. 
This seems to be consistent with the PIRG result at $n=1$, 
in which a quite independent measurement, namely,  
the jump of the double occupancy 
indicating the first-order BCMT appears 
at $U_{c}=3.25\pm0.05$~\cite{KI}. 
The metal-insulator boundary except the BCMT point is the boundary of 
the filling-control Mott-transition (FCMT). 
It should be noted that the phase boundary remarkably has a V-shaped structure rather than a U-shaped one.
As in the black dashed curve in Fig.~\ref{fig:mu-Ubulk}, the phase boundary of FCMT very close to the BCMT
appears to show a small deviation from the linear fitting, this uncertainty coming from the uncertainty of the 
size extrapolation is small. It is not clear whether the phase boundary has a round shape (U-shape structure) 
near BCMT, or V-shaped until BCMT.  However, the overall shape of the phase boundary is well represented by 
the V shape and the round structure is limited to a very narrow region if it exists.
This remarkable V-shape structure will further be discussed in \S 4.3.

\subsection{Carrier dependence of chemical potential 
and charge compressibility}

\begin{figure}
\begin{center}
\epsfxsize=8cm \epsfbox{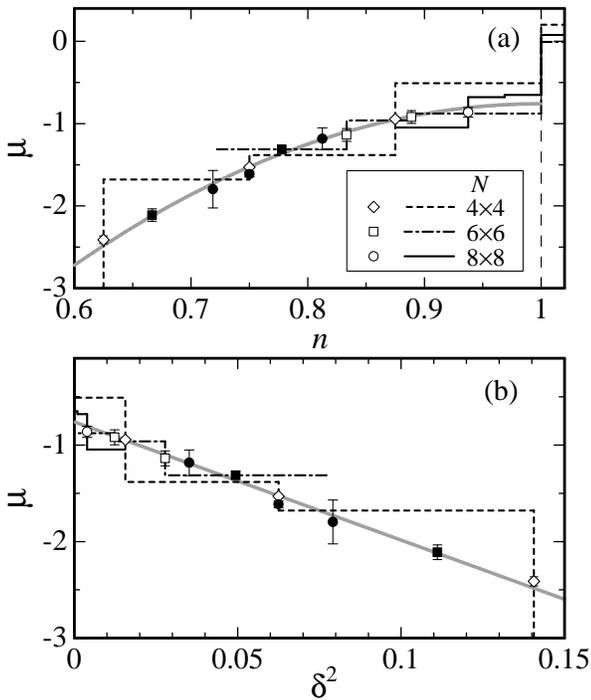}
\end{center}
\caption{
(a) Chemical potential $\mu$ vs filling $n$ for $t=1.0$ and $t'=-0.2$ 
on the square lattice. 
(b) $\mu$ vs $\delta^{2}$.
Dashed line, dash-dotted line, and solid line represent 
$N=4 \times 4$, $6 \times 6$, and $8 \times 8$ systems, 
respectively, obtained by GPIRG method. 
Open symbols are plotted at middle point of each step of $\mu$-$n$ line.
Solid symbols are obtained by PIRG method for closed shell structure.
Gray curves in both panels are least square fits.
}
\label{fig:mu-n}
\end{figure}

Though we have obtained the slope ${\delta}U/{\delta}\mu$ as in 
Fig.~\ref{fig:mu-Ubulk}, the slope itself does not tell us the order of the 
metal-insulator transition. 
In the literature, the FCMT is identified as the continuous transition with a diverging compressibility by QMC studies~\cite{FI,FI1993}, while the BCMT shows the first-order transition with a jump of the double occupancy, which has been obtained from the PIRG study~\cite{KI}.  Before discussing this contrast, we apply GPIRG method to confirm whether the continuous character of the FCMT is stable and reliable, since the GPIRG method is an independent technique from the QMC method.   
To identify the order of the transition 
we calculate the filling dependence of the chemical potential. 
In Fig.~\ref{fig:mu-n}(a) we show 
the chemical potential $\mu$ vs. filling $n=N_{e}/N$ for $U=4.0$. 
The dashed line, the dash-dotted line, and 
the solid line represent the $\mu$-$n$ lines calculated by the GPIRG 
for the $N=4 \times 4$, $6 \times 6$ and $8 \times 8$ systems, respectively. 
We plot the open symbols at the middle point of each step 
for the $N=4 \times 4$ (open diamond), $6 \times 6$ (open square), and 
$8 \times 8$ (open circle) systems. 
We also plot the data by the filled symbols 
calculated from eq.~(\ref{eq:chem}) by the PIRG in the closed-shell structure 
for the $N=4 \times 4$ (filled diamond), $6 \times 6$ (filled square), and 
$8 \times 8$ (filled circle) systems. 
As seen in Fig.~\ref{fig:mu-n}(b) 
all symbols are fitted well by the function 
\begin{eqnarray}
\mu=\mu_{c}-a\delta^2, 
\label{eq:fit}
\end{eqnarray}
with $\delta=1-n$. 
The gray lines in Fig.~\ref{fig:mu-n}(a) and Fig.~\ref{fig:mu-n}(b) 
are the least square fit with fitting function, eq.~(\ref{eq:fit}), 
which gives $a=-12.26\pm 0.27$ 
and $\mu_{c}=-0.76 \pm 0.01$. 
This indicates that the charge compressibility has the form, 
\begin{eqnarray}
\chi_{c}=\left( \frac{\partial\mu}{\partial n} \right)^{-1}
\sim \frac{1}{2a}(1-n)^{-1}
\label{eq:chi}
\end{eqnarray}
away from half filling 
as shown by Furukawa and Imada~\cite{FI}. 
If the filling $n$ becomes discontinuous as a function of $\mu$, 
the first-order transition occurs. 
Namely, unstable $n$ leads to the inhomogeneous ground state 
which has the hole-rich and -poor regimes. 
On the contrary, 
the fitting of eq.~(\ref{eq:fit}) in Fig.~\ref{fig:mu-n} 
seems to show 
$\chi_{c} \to \infty$ for $n \to 1$ and 
$\chi_{c}=0$ at $n=1$. 
In Fig.~\ref{fig:mu-n}, however, 
the closest data point to half filling is $n=0.9375$. 
Hence, we safely conclude that 
the phase separation does not occur for the carrier density 
larger than $6\%$ for $U=4.0$. 
Absence of the phase separation in the Hubbard model 
has been reported in refs.~\cite{MS,FI} 
for the no-frustration case, $t'=0$ 
as well as for a frustrated case at $t'=0.2$~\cite{FI1993}. 
Our result is perfectly consistent with the QMC result and 
$a$ and $\mu_{c}$ show quantitative agreement with the result in 
Ref.~\cite{FI1993}.
%
%

When the first-order transition occurs at the BCMT, 
one could expect that 
the phase separation also occurs at the FCMT in general. 
The numerical results discussed above, however, show rather contrasted behavior.
The BCMT shows clear first-order transition, 
while FCMT is most likely continuous with 
a diverging compressibility at the transition point.  

In the framework of the commensurate AF mean-field approximation
(Hartree-Fock approximation) (see eqs.~(\ref{eq:HFeq1}) and (\ref{eq:HFeq2})), 
the AF metal appears by the carrier doping to the AF insulator 
and the phase separation between the paramagnetic metal and 
the AF metal occurs: 
For $t'=-0.2$ and $U=4.0$, the phase-separated region appears 
for $0.69 \le n \le 0.85$ in the hole-doped case. 
Namely, the phase separation appears for the interval of filling, $16 \%$. 
The first-order BCMT occurs at $U_{c}^{\rm HF}=2.064$
at $n=1$~\cite{KonMori} 
in the mean-field approximation~\cite{LinHirsh,KonMori,vollhardt}, 
where the jump of the double occupancy is estimated at 0.04. 
By the PIRG calculation at $n=1$
the jump of the double occupancy is estimated at 
$0.04$ at $U_{c}=3.25 \pm 0.05$~\cite{KI}. 
Though the jump of the double occupancy at $n=1$ by the PIRG method is comparable to 
the mean-field approximation, 
the region of the phase separation at the FCMT 
should be largely reduced from the mean-field result 
if it existed.
This type of large phase-separation region in the mean-field approximation disappears in our
result after considering the quantum fluctuation effect by the GPIRG.  
In addition, the phase separation in the mean-field approximation occurs between a paramagnetic metal 
and antiferromagnetic metal.  The FCMT itself is a continuous transition even in the mean-field solution, which is the same as our result,
although the antiferromagnetic metal seen in the mean-field approximation is likely to become absent in metals again because of the quantum
fluctuations. 

We also comment about the contrast between the BCMT and the FCMT from viewpoint of 
the bound-state formation. 
Since the electrostatic 
attractive force works between a doublon (a site occupied by two electrons) and a holon (a site unoccupied by electrons), 
the bound state between a doublon and a holon is formed at the BCMT. 
The Mott insulating state is indeed regarded as the phase where the doublons and holons
all form bound states in pairs.
Here we recall that 
in the dislocation-vector system~\cite{YS,YS2} and 
the Coulomb-gas system~\cite{LW}, 
the Kosterlitz-Thouless transition changes to the first-order transition 
as the ratio of the binding energy of vortex cores and the core energy 
exceeds a threshold. 
The same mechanism may work in the BCMT and FCMT.
Namely, in case of the BCMT, relatively large 
attractive force between a holon and a doublon induces the first-order phase transition. 
In the case of the FCMT, however, the mechanism is quite different. 
The transition in this case is controlled by an interaction between 
two holons (or between two doublons). 
Apparently, the attraction between two holons should be much smaller than 
the attraction between a holon and a doublon. 
In the present system, 
if the attractive force works among holons (doublons) 
at the hole (electron)-doped Mott insulator, 
the bound state is formed between holons (doublons) and the first-order FCMT 
may occur. 
Our calculation  for $U=4.0$  shows that 
the binding energy between holons 
is weak enough so that the transition becomes continuous 
or at most the discontinuity is very small. 
A sharp contrast between the FCMT and BCMT is naturally understood from this 
completely different origin of the attractions.

\subsection{Thermodynamic analysis}

To understand the contrast between the BCMT and the FCMT in detail, 
we derive a relation between the slope of the 
metal-insulator-transition boundary in the $\mu$-$U$ phase diagram and 
some other thermodynamic quantities. 
Let us consider the expansion of the ground-state energy by 
$\mu$ and $U$.
\begin{eqnarray}
\lefteqn{E(\mu+\delta\mu,U+{\delta}U)}\nonumber \\ 
&=&E(\mu,U)+\left(\frac{{\partial}E}{{\partial}\mu}\right)_{U  }\delta\mu +\left(\frac{{\partial}E}{{\partial}U  }\right)_{\mu}{\delta}U \nonumber \\
& &+O((\delta\mu)^2,(\delta U)^2). 
\label{eq:E1}
\end{eqnarray}
In the $\mu$-$U$ phase diagram, along 
the metal-insulator-transition boundary, 
the energies of the metal and insulator phases may not change: 
\begin{eqnarray}
\lefteqn{E_{I}(\mu,U)-E_{M}(\mu,U)}\nonumber \\
& & = E_{I}(\mu+\delta\mu,U+{\delta}U) - E_{M}(\mu+\delta\mu,U+{\delta}U)\nonumber \\
& & = 0, \label{eq:E2}
\end{eqnarray}
where subscript I(M) represents that the expectation value 
is taken in the insulating (metallic) ground state. 
If the first-order metal-insulator transition occurs, 
the slope of the transition line in the $\mu$-$U$ phase diagram is 
determined by the ratio of the jump of filling and double occupancy. 
Namely, from eq.~(\ref{eq:E1}) and eq.~(\ref{eq:E2}) 
the following equation is derived: 
\begin{eqnarray}
\frac{{\delta}U}{{\delta}\mu}=\frac{n_{I}-n_{M}}{D_{I}-D_{M}}, 
\label{eq:derivum}
\end{eqnarray}
where
\begin{eqnarray}
\left(\frac{{\partial}E}{{\partial}\mu}\right)_{U  }&=&-Nn, 
\label{eq:E3} \\
\left(\frac{{\partial}E}{{\partial}U  }\right)_{\mu}&=&
\sum_{i=1}^{N}
\left\langle \left(n_{i\uparrow}-\frac{1}{2}\right)
\left(n_{i\downarrow}-\frac{1}{2}\right)
\right\rangle, 
\nonumber \\ 
&\equiv& ND. 
\label{eq:E4}
\end{eqnarray}
At the BCMT point ${\delta}U/{\delta}\mu=0$ 
if the first-order transition occurs, 
since the numerator of eq.~(\ref{eq:derivum}) 
becomes zero because $n_{I}$=1 and $n_{M}=1$ are both satisfied, 
but the denominator is finite. 

In the case of the continuous metal-insulator transition, 
let us consider the expansion of filling $n$:
\begin{eqnarray}
\lefteqn{n(\mu+\delta\mu,U+{\delta}U)}
\nonumber \\ 
&=&n(\mu,U)+\left(\frac{{\partial}n}{{\partial}\mu}\right)_{U  }\delta\mu
+\left(\frac{{\partial}n}{{\partial}U  }\right)_{\mu}{\delta}U
\nonumber \\
& &+O\left(({\delta}\mu)^2,({\delta}U)^2\right). 
\label{eq:n1}
\end{eqnarray}
By a parallel discussion to the above derivation, 
the slope of the metal-insulator boundary 
in the $\mu$-$U$ phase diagram is derived as
\begin{eqnarray}
\frac{{\delta}U}{{\delta}\mu}=-\frac{\chi_{c}}
{\left(\frac{{\partial}n_{M}}{{\partial}U}\right)_{\mu}}, 
\label{eq:n2} 
\end{eqnarray}
where $\chi_{c}$ is the charge compressibility 
\begin{eqnarray}
\chi_{c}=
\left(
\frac{{\partial}n_{M}}{{\partial}\mu}
\right)_{U}, 
\label{eq:ckai}
\end{eqnarray}
in the metallic phase and $({\partial}n_{I}/{\partial}\mu)_{U}=0
=({\partial}n_{I}/{\partial}U)_{\mu}$ is used.

\begin{figure}
\begin{center}
\epsfxsize=8cm \epsfbox{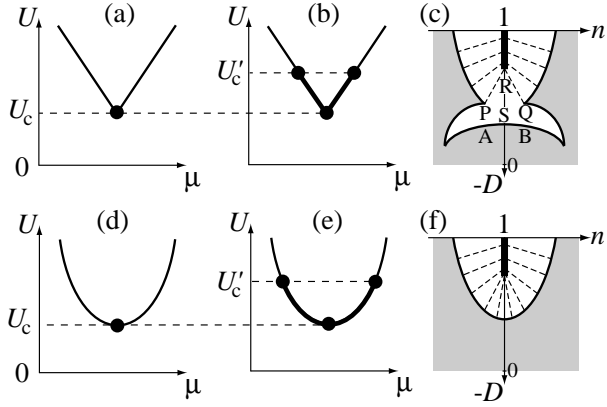} 
\end{center}
\caption{
Upper panel:
Possible schematic phase diagrams with a V-shaped metal-insulator boundary,
where the first order transitions are indicated by solid circles or thick
lines
while thin lines represent continuous transitions.
 (a) shows the first-order transition only at the corner while continuous
in the edges , and (b) shows the case with the first order transition both
at the corner and edges (in the
part drawn by thick lines).
(c) $-D$-$n$ phase diagram corresponding to the case (b).
Lower panel: 
Possible schematic phase diagrams of U-shaped metal-insulator
 boundary with the same notations as the V-shaped cases.
(d) shows the case with the first-order transition only at the bottom, and
(e) shows the case with the first-order transition both at the bottom and
edges (in the part drawn by thick lines).
(f) $-D$-$n$ phase diagram
corresponding to (e).
In (c) and (f),  the insulator phase is represented by thick line
at $n=1$, and the metallic phase is represented by the gray area.
$-D=0.25$ is taken along the $n$ axis. 
The miscibility gap due to the phase separation is given by the white area
with dashed curves,
which connect the coexisting metallic and insulating points.  Namely, the
two ends of a dashed curve  represent the jump of $D$ and $n$ at the
first-order transitions for a given $U<U'_{c}$.  
$U_{c}$ denotes the BCMT point
and
$U'_{c}$ denotes the possible critical point where
the first-order transition converts to
the continuous transition.
The case (b) (and (c)) is unlikely to be realized. 
The present numerical result supports the case (a) (see the text). 
}
\label{fig:UV_shape}
\end{figure}

From the relations eq.~(\ref{eq:derivum}) and eq.~(\ref{eq:n2}), 
it turns out that the V-shaped structure implies a quite different character 
between the bandwidth- and filling-control transitions.  
In fact, if one assumes the first-order transition for the bandwidth control 
at the corner of the V shape as we indeed observed, $dU/d\mu$ is not well defined at the corner
(see $U_{c}$ in Fig.~\ref{fig:UV_shape}(a)).  
At this corner, we have observed a jump in $D$, namely a nonzero $D_{I}-D_{M}$ 
while the filling should not show a jump at this corner.  
This implies $dU/d\mu=0$, which does not contradict the character of the corner.  

However, if the first-order character is also retained in the edge of the phase boundary 
as drawn by thick lines in Fig.~\ref{fig:UV_shape}(b), 
namely for the filling-control transition, the V shape together with eq.~(\ref{eq:derivum}) 
results in an emergence of the abrupt jump in $n_{I}-n_{M}$ 
in the region above $U_c$ but infinitesimally close to the corner, which is unlikely.  
This is also understood in Fig.~\ref{fig:UV_shape}(c) 
which is the $D$-$n$ phase diagram corresponding to 
Fig.~\ref{fig:UV_shape}(b). 
The abrupt jump in $n_{I}-n_{M}$ at $U$ infinitesimally close to $U_{c}$ 
in Fig.~\ref{fig:UV_shape}(b) leads to a nonzero, finite phase-separated region at $U \sim U_{c}$ 
in the $D$-$n$ phase diagram as in the coexistence of R and Q (or R and P)
in Fig.13(c). 
At $U=U_c$ the ground-state energy at P, R and Q are degenerate also with 
that at S, because of the first-order nature of the BCMT. 
Namely, 
the double-well structure in the energy as a function of $n$ (local minima
at $n_{I}$ and $n_{M}$)
forces us to allow three different metal phases separated by first-order
transitions as shown as P, Q and S in Fig.~\ref{fig:UV_shape}(c).
Then at $U_c$, a metal with the density at P or Q is realized at the same chemical potential as the point R and S. Now the same density of metal as P and Q has to be realized also at A and B when the chemical potential is shifted from S, because the points A and B are realized from S by changing the chemical potential adiabatically.  Then at the same $U$, the same density of metal (namely P and A ( as well as Q and B) ) is realized at different two chemical potentials, which is unphysical. 
Therefore the case 
Fig.~\ref{fig:UV_shape}(b) is unlikely to be realized.
The above discussion concludes that, 
{\it 
the V-shaped metal-insulator transition lines
together with the first-order BCMT is not compatible with the presence of the first-order FCMT near the BCMT.
}

If the transition converts to the continuous character 
for $U$ infinitesimally close to $U_{c}$ in Fig.~\ref{fig:UV_shape}(a), 
it implies a diverging compressibility 
and diverging $dn/dU$ with a finite ratio for FCMT.  This is certainly possible.  
 Therefore, the V-shaped structure is consistent with the abrupt conversion 
 of the first-order character observed in the bandwidth-control transition to the continuous character 
 of the filling-control transition supported from the data in the previous subsection.  

In the case of the U-shaped structure, 
it is possible that the first-order transition occurs at the bottom $U_{c}$ 
as in Fig.~\ref{fig:UV_shape}(d), 
and also that the first-order transition is retained until 
the critical points $U'_{c} (> U_{c})$ 
as drawn by thick lines in Fig.~\ref{fig:UV_shape}(e). 
The latter is because 
$dU/d{\mu}$ changes from zero to finite values continuously as $U$ changes from $U_{c}$. 
Namely, 
as $U$ changes from $U_{c}$ to $U'_{c}$, 
$D_{I}-D_{M}$ changes from a finite value to zero,   
and $n_{I}-n_{M}$ changes from zero to finite values and finally becomes zero, continuously. 
%
%
In the corresponding $D$-$n$ phase diagram, Fig.~\ref{fig:UV_shape}(f) 
with the same notations as Fig.~\ref{fig:UV_shape}(c), 
we see that there exists no unphysical phase separation within the metallic phase 
in contrast with Fig.~\ref{fig:UV_shape}(c). 
In this case, 
the phase-separated region terminates continuously at both ends of $U'_{c}$ and $U_{c}$, 
and the insulator phase is realized for $U>U_{c}$. 
This argument does not in principle exclude the possibility that ${U'}_{c}$ becomes infinity. 

In our calculation, 
within the numerical accuracy, we do not exclude the possibility that the V shape could in reality have a U shape corner structure in the very vicinity of the BCMT
 with $dU/d\mu$ being well defined everywhere.  In this case, we do not need to assume an unphysical jump 
 for the filling control as mentioned above.  
 Even though, a rather sharp change observed at the ``corner" of the phase boundary supports a rapid change 
 of the character between two routes of the transitions.

%
%

We are now left with the possibilities of (a), (d), and (e) in Fig.~\ref{fig:UV_shape},
where the round part is limited to the region close to the corner if
(d) and (e) apply, because the overall V shape is clear.
In fact, the analysis of the charge gap in the next subsection more precisely shows that the 
gap opens linearly with $U-U_{c}$.  Therefore the phase diagram is most likely to belong to the class (a) 
in Fig.~\ref{fig:UV_shape}, the FCMT is continuous while BCMT is the first-order transition.
We also analyze whether the first-order transition can be consistent with the slope in the V-shaped
structure away from the corner.  The slope $\delta U/\delta \mu$ is roughly 2.4 in the hole-doped 
part. When the first-order transition occurs, and the jump of the double occupancy
has a similar value to the jump at BCMT ($\sim 0.04$), the phase separation should extend to the 
doping concentration $\sim 0.1$.  However, this is clearly not the case from our numerical results, 
and the phase separation is limited at most to the region less than 0.06.  
Rather sharp difference in the character of FCMT from BCMT becomes recognizable also from such analyses. 

In the $\Upsilon$-shaped structure case, such that 
the metal-insulator transition line behaves as 
$\mu \sim (U-U_{c})^{\gamma}$ with $\gamma>1$, 
the situation is the same as the V-shaped case. 
The essential singular form such as 
$\mu \sim \exp(-2\pi\sqrt{t/U})$ 
is known to be realized for $t'=0$ in the HMSL~\cite{hirsch1985b} 
with the metal-insulator transition at $U=0$. 
We are unaware of the type of transitions in a concrete model except the case $U_{c}=0$.
If the system with this form with the first-order BCMT point $U_{c}>0$ exists, 
it is classified to the same class as discussed above. 
Therefore, the first-order transition is likely to occur only at the corner.

\subsection{Charge excitation gap}

The charge gap is defined by 
\begin{eqnarray}
\Delta_{c}\equiv \frac{1}{2} \left[
\mu\left(
\frac{N+1}{N}
\right)
-\mu
\left(
\frac{N-1}{N}
\right)
\right], 
\nonumber
\end{eqnarray}
where 
$
\mu((2N'-1)/N)=\{E(N',N')-E(N'-1,N'-1)\}/2
$
and 
$E(N_{\uparrow},N_{\downarrow})$ is the ground-state 
energy with number of 
up(down) spin, $N_{\uparrow}(N_{\downarrow})$. 

\begin{figure}
\begin{center}
\epsfxsize=8cm \epsfbox{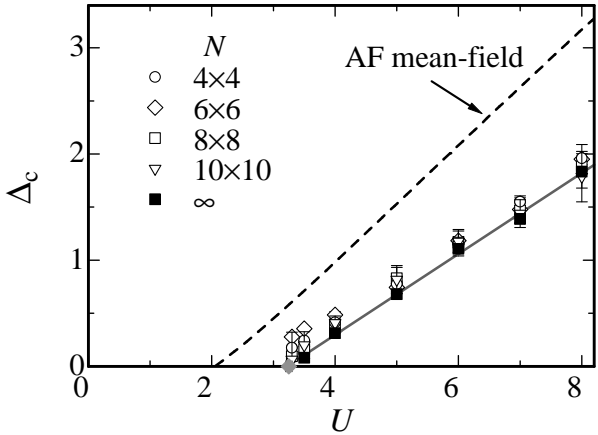} 
\end{center}
\caption{
$U$-dependence of the charge gap 
for $t=1.0$ and $t'=-0.2$ 
in 
$N=4 \times 4$ (open circle), $6 \times 6$ (open diamond), $8 \times 8$ 
(open square), and $10 \times 10$ (open triangle) systems. 
Solid squares represent the charge gap for $N \to \infty$.
Solid line is least square fit of $\Delta_{c}
(N \to \infty)$ (see text). 
Gray diamond is located at $U_{c}=3.25 \pm 0.05$ determined 
by the jump of the double occupancy at $n=1$ by PIRG~\cite{KI}.
Dashed line represents the charge gap $\Delta_{c}^{\rm HF}$ 
by the AF mean-field approximation. 
}
\label{fig:cgap}
\end{figure}

Figure~\ref{fig:cgap} shows the charge gap calculated by the GPIRG method.
Solid squares represent the charge gap, $\Delta_{c}$ 
in the thermodynamic limit. Note that, in the GPIRG method, the charge gap is determined by
the distance in $\mu$ between two phase boundaries in Fig.~\ref{fig:mu-Ubulk}. 
For the extrapolation, we have used the scaling function 
\begin{eqnarray}
\Delta_{c}(N)=\Delta_{c}+\frac{\Delta'}{\sqrt{N}},  
\nonumber
\end{eqnarray}
as in the Hartree-Fock AF gap equation~\cite{FI}. 
Since the opening of the gap $\Delta_c$ is very well described by a linear $U$ dependence, 
the solid line is fitted by the linear line of $\Delta_{c}(N \to \infty)$ 
for $U=3.5, 4.0, 5.0, 6.0, 7.0$ and 8.0. 
>From this fit, the critical value $U_{c}$ is estimated at $U_{c}=3.23$. In Fig.~~\ref{fig:cgap}, 
the gray diamond represents 
$U_{c}=3.25 \pm 0.25$ at which 
the jump of the double occupancy appears at $n=1$ 
in the PIRG calculation~\cite{KI}. 
These two independent estimates agree each other excellently. 
The charge gap grows from 0 continuously even at the first-order metal-insulator transition 
point $U_{c}$ and shows remarkably a linear dependence on $U$.
This continuous growth of the charge gap from zero is rather obvious because the 
first-order transition indicates a level crossing of the metallic and insulating states,
and the energy difference between the metallic and insulating states increase continuously 
from zero after the level crossing, while the charge gap in the insulating side must be smaller 
than this difference.

In the Hartree-Fock approximation~\cite{LinHirsh,KonMori,vollhardt}, 
the AF bands are given by 
\begin{eqnarray}
\tilde{\varepsilon}^{\pm}({\bf k})
&=&\frac{1}{2}\left[
\varepsilon({\bf k})+\varepsilon({\bf k'}) 
\right.
\nonumber
\\
&\pm&
\left.
\{\varepsilon({\bf k})-\varepsilon({\bf k'}) \}
\sqrt{1+\frac{U^2 m^2}{\{\varepsilon({\bf k})-\varepsilon({\bf k'}) \}^2}}
\right], 
\nonumber
\end{eqnarray}
where $\langle n_{i\sigma}\rangle = (n+\sigma (-1)^{|i|}m)/2$ 
and 
${\bf k}'={\bf k}+{\bf Q}$ with ${\bf Q}=(\pi,\pi)$. 
The self-consistency equations are 
\begin{eqnarray}
n &=& \frac{2}{N}\sum_{\bf k}\left\{
f(\tilde{\varepsilon}^{+}({\bf k}-\mu))
+f(\tilde{\varepsilon}^{-}({\bf k}-\mu))
\right\}, 
\label{eq:HFeq1}
\\
1&=&\frac{2U}{N}\sum_{\bf k}
\frac{
 f(\tilde{\varepsilon}^{+}({\bf k})-\mu) 
-f(\tilde{\varepsilon}^{-}({\bf k})-\mu)
}
{
\sqrt{
{\{\varepsilon({\bf k})-\varepsilon({\bf k'})
 \}^2
+U^2 m^2
}}
},
\label{eq:HFeq2}
\end{eqnarray}
where the summation for $\bf k$ is taken in the 1st-Brillouin zone 
and $f(\varepsilon)$ is the Fermi distribution function. 
For $t'=-0.2$ 
the indirect gap appears between the AF bands. 
Namely, the charge gap $\Delta^{\rm HF}_{c}$ 
is calculated from the energy difference between the bottom of the 
upper band 
$\tilde{\varepsilon}^{-}(\pi/2,\pi/2)$
and the top of the lower band 
$\tilde{\varepsilon}^{+}(\pi,0)$ 
by using the solution of eqs.~(\ref{eq:HFeq1}) and 
(\ref{eq:HFeq2}), $\mu$ and $m$, for $n=1$:
\begin{eqnarray}
\Delta^{\rm HF}_{c}\equiv
\frac{1}{2}
\left[
\tilde{\varepsilon}^{-}\left(
\frac{\pi}{2},\frac{\pi}{2}
\right)
-
\tilde{\varepsilon}^{+}
\left(
\pi,0
\right)
\right]. 
\nonumber
\end{eqnarray}
The charge gap $\Delta^{\rm HF}_{c}$ is shown by 
the dashed line in Fig.~\ref{fig:cgap}. 
Here, the order parameter shows the jump from $m=0$ to $0.391$
at the first-order-transition point 
$U_{c}^{\rm HF}$=2.064~\cite{KonMori}, but 
the charge gap appears from 0 to finite values continuously 
for  $U \ge U_{c}^{\rm HF}$. 
Comparing $\Delta^{\rm HF}_{c}$ with 
$\Delta_{c}$ obtained by the GPIRG method, 
we see that $U_{c}$ is larger than $U_{c}^{\rm HF}$ 
and $\Delta_{c}$ is reduced from 
$\Delta^{\rm HF}_{c}$ by the electron-correlation effect: 
for example, 
$\Delta_{c}/\Delta^{\rm HF}_{c}=0.32$ at $U=4.0$ and 
$0.58$ at $U=8.0$. 

It is interesting to estimate the parameters of the HMSL as 
the effective model for the cuprates: 
Optical conductivity data of $\rm La_2CuO_4$~\cite{uchida}  
indicate the charge gap $2\Delta_{c}$ of the order of $1.5 \sim 2$ eV. 
If the HMSL describes low energy physics of the material, 
the NN hopping is estimated at $t=0.43$ eV 
according to ref.~\cite{expparam,Andersen}. 
Hence, the value of $U/t$ is estimated at $7.8\sim 9.3$ 
from Fig.~\ref{fig:cgap}. 

%
%

Since the exponent of the charge gap $\alpha$ defined by $\Delta_{c}\propto [(U-U_{c})/t]^{\alpha}$ 
is consistent with $\alpha=1$ both in the Hubbard model in two dimensions and in the Hartree-Fock result,
it may be a universal feature irrespective of the dimensionality, except a special case of $U_{c}=0$ in the perfectly nested lattice.
This universality of the exponent is intriguing because the linear opening of the gap also means 
the V-shaped phase boundary and supports the 
first-order transition for the BCMT coexisting with the continuous transition for the FCMT.   
The opening of the Mott insulating gap as a function of $(U-U_{c})/t$ may be compared with
the experimental results in the literature.  In the perovskite compounds, R$_{1-x}$Ca$_x$TiO$_3$,
the opening of the charge gap has been studied by Katsufuji, Okimoto and Tokura~\cite{Katsufuji,IFT}. Here R represents rare earth ions
to control effective $(U-U_{c})/t$.  The optical gap as well as the 
activation energy in the resistivity show an opening of the gap well scaled by a linear function of 
$(U-U_{c})/t$.  Our calculated result is quite consistent with this observation.
Although a narrow doping region of the doped system in real three-dimensional perovskite may be under 
influence of impurity localization and also of the antiferromagnetic-metal phase, it turns out that 
the linear opening of the charge gap is tightly connected with a singularly different characters between
FCMT and BCMT.  
When the critical point of the Mott transition characterized by the compressibility 
divergence occurs close to zero temperature as supported from the continuous character of FCMT, the character 
of the Mott transition has to be analyzed not as the classical Ising-type transition~\cite{Kotliar} but as the quantum phase transition.  If the critical point is normally characterized by the 
vanishing $k^2$ dispersion replaced with the dominant $k^4$ dispersion, this dynamical exponent $z=4$ leads to
the compressibility as $\kappa \propto \delta^{1-z/d}$ with the doping concentration $\delta$ in $d$- spatial dimensions in the scaling analysis~\cite{Imada1995}. In two dimensions, this leads to $\kappa \propto \delta^{-1}$, which is consistent with the numerical as well as experimental results~\cite{IFT}.  In three dimensions, this leads to $\kappa \propto \delta^{-1/3}$.
The divergence is much weaker than the two-dimensional case
and rather difficult to see the compressibility divergence experimentally as compared to two-dimensional systems.  This is consistent again with the experimental indications~\cite{Fujimori}, although the surface effect in the photoemission experiments has to be carefully considered.

\section{Summary} 
We have studied the ground-state properties of the 
Hubbard model on the square lattice (HMSL) 
with nearest-neighbor and next-nearest-neighbor hoppings. 
The computation by our new algorithm has made it possible to draw a phase diagram of the Hubbard model in the parameter space of the interaction $U$, chemical potential $\mu$ and the band structure (or amplitude of frustration). 

Our new method is summarized as follows: To study the bandwidth-control Mott transition and the filling-control-Mott transition by a unified approach, we have developed 
an algorithm of the grand-canonical path-integral renormalization group 
(GPIRG) method. 
To treat the system in the grand canonical ensemble, 
the particle-hole transformation is applied to the Hamiltonian and the basis states. 
To reach the ground state for the chemical potential $\mu$, 
the Stratonovich-Hubbard transformation which hybridizes up and down 
electrons is introduced.
To avoid the trapping in a local minimum with a specific electron number, 
it is efficient to use the pseudo-chemical potentials (fictitious chemical potential)
in the kinetic-term projection, 
which are different from the original chemical potential $\mu$. 
This algorithm does not suffer from the negative-sign problem, which often 
becomes serious in the quantum Monte Carlo method 
and can be applied to any Hamiltonian in any lattice structure and dimension 
with arbitrary boundary conditions.
The GPIRG results in the ground-state energy and the physical quantities 
such as the equal-time spin correlation and the momentum distribution function 
show good agreement with the results by the 
exact diagonalization and the quantum Monte Carlo methods. 
The GPIRG is shown to be useful to calculate the chemical-potential dependence 
of physical quantities. 

Our calculated results by the GPIRG for the two-dimensional Hubbard model is summarized as follows:
The ground-state phase diagram of the HMSL 
in the plane of 
the chemical potential $\mu$ and the interaction $U$  
is precisely determined.
The contrast between the bandwidth and filling-control 
Mott transitions clarified in the literature are more firmly confirmed on the basis of 
a unified approach here, in the following fashion:
The carrier-density dependence on the chemical potential 
indicates that 
the phase separation does not occur (for example, at least for carrier doping larger than $6\%$ 
at $U=4.0$), implying the continuous character of the transition with enhanced fluctuations near the transition as seen in the diverging charge compressibility. In sharp contrast, the bandwidth-control Mott transition shows a clear first-order character with a jump of the double occupancy (for example at $U_{c}=3.25\pm0.05$ for $t'=-0.2$). 

The phase boundary of metals and the Mott insulator shows that the charge gap $\Delta_c$ opens at $U=U_{c}$ and 
shows marked linear dependence 
on $U$ for $U>U_{c}$, namely $\Delta_c \sim \alpha (U-U_{c})$.
The resultant V-shaped singular phase boundary in the plane of $U/t$ and $\mu$ (as seen in Fig.11) is shown to be consistent with the sharp contrast between the continuous filling-control transition and the first-order character of the bandwidth-control transition.   
A general relation of the slope of the metal-insulator transition line 
in the $\mu$-$U$ phase diagram and physical quantities are also derived: 
In the case of the first-order metal-insulator transition, 
$\delta U/\delta \mu$ is expressed by the ratio of the jumps in the filling and 
the double occupancy $\Delta n/\Delta D$. 
In the case of the continuous transition, 
$\delta U/\delta \mu$ is expressed by the ratio of 
the compressibility and $dn/dU$ in the 
metallic phase. These relations support that the V-shaped phase boundary is resulted from the first-order BCMT coexisting with
continuous FCMT with diverging compressibility.
Experimental results of Ti perovskite compounds are favorably compared with this V shape and the contrast between BCMT and FCMT.

\noindent
{\bf Acknowledgments}

One of the authors (S. W.) 
would like to thank H. Yokoyama and H. Kohno 
for valuable discussions.
The work is supported by Grant-in-Aid for young scientists, No. 15740203 
from the Ministry of Education, Culture, Sports, Science and Technology, 
Japan. 
A part of our computation has been done at the supercomputer center 
in the Institute for Solid State Physics, University of Tokyo.

\appendix
\section{BCS wavefunctions in the GPIRG framework}
In this appendix, we show the BCS wavefunction 
in the GPIRG framework and make some comments.

The BCS wavefunction is defined by 
\begin{eqnarray}
|\psi_{\rm BCS}\rangle= 
\prod_{k}
\left(
u_{k}+
v_{k}c_{k\uparrow}^{\dagger}c_{-k\downarrow}^{\dagger}
\right)
|0\rangle,
\label{eq:BCSwf}
\end{eqnarray}
where $u_{k}$ and $v_{k}$ satisfy $u_{k}^{2}+|v_{k}|^{2}=1$.

By the canonical transformation of eq.~(\ref{eq:PHtrans}), 
the BCS wavefunction is transformed as follows~\cite{YokoShiba}:
\begin{eqnarray}
|\psi_{\rm BCS}\rangle &=& 
\prod_{k}
\left(
u_{k}+v_{k}c_{k}^{\dagger} d_{k}
\right)
\prod_{k'}
d_{k'}^{\dagger}|0\rangle
\nonumber \\
&=& 
\prod_{k}
\left(
u_{k}d_{k}^{\dagger}+v_{k}c_{k}^{\dagger}
\right)|0\rangle
\end{eqnarray}

The BCS wavefunction is given as the $2N \times N$ Slater matrix: 
\[
[\phi]_{jk}= \left\{
\begin{array}{ll}
u_{k}{\rm e}^{{\rm i}{\bf k}\cdot{{\bf R}_{j}}}, & 
{\rm for}~j=1,...,N \\
v_{k}{\rm e}^{{\rm i}{\bf k}\cdot{{\bf R}_{j}}}, &
{\rm for}~j=N+1,...,2N,  
\end{array}\right.
\]
where $k$ specifies $N$ points in the Brillouine zone. 
Here $u_{k}$ and $v_{k}$ have the following forms:
\begin{eqnarray}
u_{k}&=&\frac{1}{\sqrt{2}}
\left(
1+\frac{\xi_{k}}{\sqrt{\xi_{k}^{2}+|\Delta_{k}|^{2}}}
\right)^{1/2}, 
\nonumber \\
v_{k}&=&\frac{\rm e^{{\rm i}\varphi}}{\sqrt{2}}
\left(
1-\frac{\xi_{k}}{\sqrt{\xi_{k}^{2}+|\Delta_{k}|^{2}}}
\right)^{1/2},
\nonumber
\end{eqnarray}
where $\xi_{k}=\varepsilon_{k}-\mu$ with 
$\varepsilon_{k}=-2t[\cos(k_{x})+\cos(k_{y})]+4t'\cos(k_{x})\cos(k_{y})$
and $\mu$ is the chemical potential. 
The gap function can be given for several symmetries: For example, 
\[
\Delta_{k} = \Delta \times \left\{
\begin{array}{ll}
1, & \quad {\mbox {\rm isotropic~s~wave}} \\
(\cos(k_{x})+\cos(k_{y})), & \quad {\mbox {\rm s~wave~(cross)}} \\
(\cos(k_{x})-\cos(k_{y})), & \quad {\mbox d_{x^{2}-y^{2}}~{\rm wave}} \\
2\cos(k_{x})\cos(k_{y}), & \quad {\mbox {\rm s~wave~(diagonal)}} \\
2\sin(k_{x})\sin(k_{y}). & \quad {\mbox d_{xy}~{\rm wave}} 
\end{array}\right. 
\]
By taking the superposition of the columns of $[\phi]_{jk}$, 
$[\phi]_{jk}$ can be transformed into the real one: 
For example, in case that two columns are specified by $\bf k$ and $-\bf k$, 
they are expressed by 
$\cos({\bf k}\cdot{{\bf R}_{j}})$ 
and $\sin({\bf k}\cdot{{\bf R}_{j}})$ by using the crystal inversion symmetry. 

In the Hubbard model with on-site Coulomb repulsion, 
we note that there is no BCS-type mean-field solution.
Hence, the most optimal state for $L=1$ is not provided by 
the Slater determinant of eq.~(\ref{eq:BCSwf}). 

\section{A proof of the 
Stratnovich-Hubbard transformation which hybridizes $c$ and $d$ particles}
In this appendix, a proof of the Stratnovich-Hubbard transformation which hybridizes $c$ and $d$ particles, 
eq.~(\ref{eq:SH}) is given.

Let us rewrite eq.~(\ref{eq:SH}) with the site index omitted: 
\begin{eqnarray}
\exp\left[-{\Delta_{\tau}}U\left\{
\frac{1}{2}\left(c^{\dagger}c+d^{\dagger}d\right)
-c^{\dagger}cd^{\dagger}d
\right\}\right]
\nonumber \\
=\frac{1}{2}\sum_{s=\pm 1} 
\exp\left[{\rm i}{\beta}s
\left(
c^{\dagger}d+d^{\dagger}c
\right)
\right], 
\label{eq:SH_A}
\end{eqnarray}
with $\beta=\cos^{-1}(\exp(-{\Delta_{\tau} U}/{2}))$ for $\Delta_{\tau} U>0$. 

The left-hand site of eq.~(\ref{eq:SH_A}) is expanded by noting 
that 
$c^{\dagger}c$, $d^{\dagger}d$, and $c^{\dagger}cd^{\dagger}d$ commute each other, 
as follows: 
\begin{eqnarray}
& &
\exp\left[-{\Delta_{\tau}}U\left\{
\frac{1}{2}\left(c^{\dagger}c+d^{\dagger}d\right)
-c^{\dagger}cd^{\dagger}d
\right\}\right], 
\nonumber \\ 
&=&\exp\left(-\frac{\Delta_{\tau} U}{2}c^{\dagger}c\right)
\exp\left(-\frac{\Delta_{\tau} U}{2}d^{\dagger}d\right)
\exp\left(\Delta_{\tau} U c^{\dagger}cd^{\dagger}d\right), 
\nonumber \\ 
&=&\left\{
1+({\rm e}^{-\Delta_{\tau} U/2}-1)c^{\dagger}c
\right\}
\left\{
1+({\rm e}^{-\Delta_{\tau} U/2}-1)d^{\dagger}d
\right\} 
\nonumber \\ 
&\times&
\left\{
1+({\rm e}^{\Delta_{\tau} U}-1)c^{\dagger}c d^{\dagger}d
\right\}, 
\nonumber \\ 
&=&
1+({\rm e}^{-\Delta_{\tau} U/2}-1)
\left(
c^{\dagger}c+d^{\dagger}d-2c^{\dagger}c d^{\dagger}d
\right). 
\label{eq:LHS_A}
\end{eqnarray}
The right hand site of eq.~(\ref{eq:SH_A}) is expanded by noting 
the fact that 
$(c^{\dagger}d+d^{\dagger}c)^{2n}=
c^{\dagger}c+d^{\dagger}d-2c^{\dagger}cd^{\dagger}d
$
and 
$(c^{\dagger}d+d^{\dagger}c)^{2n+1}=
c^{\dagger}d+d^{\dagger}c
$ 
for $n=1,2,...,\infty$, 
as follows: 
\begin{eqnarray}
& &
\exp\left[{\rm i}{\beta}s
\left(
c^{\dagger}d+d^{\dagger}c
\right)
\right]
\nonumber
\\
&=&
1+\sum_{n=1}^{\infty}\frac{({\rm i}{\beta}s)^{n}}{n!}
\left(
c^{\dagger}d+d^{\dagger}c
\right)^{n}, 
\nonumber
\\
&=&1+{\rm i}\sum_{n=0}^{\infty}
(-1)^{n}\frac{(\beta s)^{2n+1}}{(2n+1)!}
\left(
c^{\dagger}d+d^{\dagger}c
\right), 
\nonumber
\\
& &+\sum_{n=1}^{\infty}(-1)^{n}\frac{(\beta s)^{2n}}{(2n)!}
\left(
c^{\dagger}c+d^{\dagger}d-2c^{\dagger}cd^{\dagger}d
\right), 
\nonumber
\\
&=&
1+{\rm i}\sin(\beta s)
\left(
c^{\dagger}d+d^{\dagger}c
\right)
\nonumber
\\
& &+\left\{
\cos(\beta s)-1
\right\}
\left(
c^{\dagger}c+d^{\dagger}d-2c^{\dagger}cd^{\dagger}d
\right). 
\nonumber
\end{eqnarray}
Then, after the summation of the Stratnovich-Hubbard variables $s=1$ and $s=-1$, 
we obtain: 
\begin{eqnarray}
& &
\frac{1}{2}\sum_{s= \pm 1}
\exp
\left[
{\rm i}\beta s 
\left(c^{\dagger}c+d^{\dagger}d\right)
\right]
\nonumber \\
&=&
1+
\left(
\cos\beta-1
\right)
\left(
c^{\dagger}c+d^{\dagger}d-2c^{\dagger}cd^{\dagger}d
\right). 
\label{eq:RHS_A}
\end{eqnarray}
To make eq.~(\ref{eq:LHS_A}) and eq.~(\ref{eq:RHS_A}) equal, the relation 
\begin{eqnarray}
\cos\beta-1=\exp
\left(-\frac{\Delta_{\tau} U}{2}
\right)-1, 
\nonumber
\end{eqnarray}
should be satisfied for $\Delta_{\tau} U>0$. 
This determines the value of $\beta$ as $\beta=\cos^{-1}(\exp(-{\Delta_{\tau} U}/{2}))$. 
By substituting this $\beta$ to eq.~(\ref{eq:LHS_A}) and eq.~(\ref{eq:RHS_A}), 
we finally obtain eq~(\ref{eq:SH_A}).



\begin{thebibliography}{99}
%
\bibitem{Mott} N. F. Mott and R. Peierls: Proc. Phys. Soc. London. 
A{\bf 49} (1937) 72. 
\bibitem{IFT} For review, see M. Imada, A. Fujimori and Y. Tokura: 
Rev. Mod. Phys. {\bf 70} (1998) 1039.
\bibitem{MWI}H. Morita, S. Watanabe and M. Imada: J. Phys. Soc. Jpn. 
{\bf 71} (2001) 2109 and see references therein.
\bibitem{Ishida}K. Ishida, M. Morishita, K. Yawata and H. Fukuyama: 
Phys. Rev. Lett. {\bf 79} (2002) 3451.
\bibitem{Saunders}A. Casey, H. Petel, J. Nyeki, B. P. Cowan and J. Saunders: Phys. Rev. Lett. {\bf 90} (2003) 115301.
\bibitem{IH} M. Imada and Y. Hatsugai: J. Phys. Soc. Jpn. {\bf 58} 
(1989) 3752. 
\bibitem{FI} N. Furukawa and M. Imada: J. Phys. Soc. Jpn. {\bf 61} 
(1992) 3331. 
\bibitem{FI1993} N. Furukawa and M. Imada: J. Phys. Soc. {\bf 62} (1993) 2257. 
\bibitem{KI} T. Kashima and M. Imada: J. Phys. Soc. Jpn. {\bf 70} 
(2001)  3052.
%
\bibitem{anderson} P. Fazekas and P. W. Anderson: Philos. Mag. 
{\bf 30} (1974) 423. 
\bibitem{expparam} M. S. Hybertsen, E. B. Stechel, M. Schluter, and 
D. R. Jennison: Phys. Rev. B {\bf 41} (1990) 11068. 
\bibitem{Andersen} E. Pavarini, I. Dasgupta, T. Saha-Dasgupta, 
O. Jepsen, and O. K. Andersen: 
Phys. Rev. Lett. {\bf 87} (2001) 47003. 
\bibitem{PIRG1} M. Imada and T. Kashima: J. Phys. Soc. Jpn. {\bf 69} 
(2000)  2723.
\bibitem{PIRG2} T. Kashima and M. Imada: J. Phys. Soc. Jpn. {\bf 70} 
(2001)  2287.
%
\bibitem{YokoShiba} H. Yokoyama and H. Shiba: J. Phys. Soc. Jpn. {\bf 57} 
(1987) 2482.
\bibitem{Hirsh} J. E. Hirsch: Phys. Rev. B {\bf 28} (1983) 4059.
\bibitem{Sorella2} S. Sorella: preprint(cond-mat/0009149). 
%
\bibitem{MI}T. Mizusaki and M. Imada: preprint(cond-mat/0311005).
%
\bibitem{hirsch1985b} J. E. Hirsch: Phys. Rev. B {\bf 31} (1985) 4403.
\bibitem{MS} A. Moreo and D. Scalapino: Phys. Rev. B {\bf 43} (1991) 11442. 
\bibitem{KonMori} H. Kondo and T. Moriya: J. Phys. Soc. Jpn. {\bf 65} (1996) 2559.
\bibitem{LinHirsh} H. Q. Lin and J. E. Hirsch: Phys. Rev. B {\bf 35} (1987) 3359. 
\bibitem{vollhardt} W. Hofstetter and D. Vollhardt: Ann. Physik {\bf 7} (1998) 48.
\bibitem{YS} Y. Saito: Phys. Rev. Lett. {\bf 48} (1982) 1114. 
\bibitem{YS2} Y. Saito: Phys. Rev. B {\bf 26} (1982) 6239. 
\bibitem{LW} J. Lidmar and M. Wallin: Phys. Rev. B {\bf 55} (1997) 522.
%
\bibitem{uchida} S. Uchida, T. Ido, H. Takagi, T. Arima, Y. Tokura 
and S. Tajima: Phys. Rev. B {\bf 43} (1991) 7942. 
\bibitem{Katsufuji} T. Katsufuji, Y. Okimoto and Y. Tokura: Phys. Rev. Lett.
{\bf 75} (1995) 3497. 
\bibitem{Kotliar} G. Kotliar: Eur. J. Phys. B {\bf 11} (1999) 27. 
\bibitem{Imada1995} M. Imada: J. Phys. Soc. Jpn.
{\bf 64} (1995) 2954. 
\bibitem{Fujimori} A. Fujimori: private communication.
%
\end{thebibliography}
\end{document}